\documentclass[article]{IEEEtran}

\usepackage[noadjust]{cite}
\usepackage[pdftex]{graphicx}
\usepackage[cmex10]{amsmath, mathtools}
\usepackage{amsfonts, amssymb, amsthm}
\usepackage{array}
\usepackage[usenames,dvipsnames]{color}

\theoremstyle{definition}\newtheorem{Rem}{Remark}
\theoremstyle{definition}
\theoremstyle{definition}
\theoremstyle{definition} 
\theoremstyle{definition}
 \theoremstyle{definition}
\newtheorem{Prop}{Proposition} \theoremstyle{definition}

\theoremstyle{definition}

\begin{document}

\title{Search Optimization for Minimum Load under Detection Performance Constraints \\ in Multifunction Radars}

\author{Dae-Sung~Jang,~\IEEEmembership{Student~Member,~IEEE,}
        Han-Lim~Choi,~\IEEEmembership{Member,~IEEE,}
        and~Ji-Eun~Roh,~\IEEEmembership{Member,~IEEE}
\thanks{D.S. Jang and H.-L. Choi are with the Department of Aerospace Engineering, KAIST, Daejeon, 305-701, Republic of Korea. E-mails: (dsjang@lics.kaist.ac.kr, hanlimc@kaist.ac.kr).}
\thanks{J.E. Roh is with the Agency for Defence Development, Daejeon 305-152, Republic of Korea. E-mail: (jeroh@add.re.kr ).}
\thanks{All correspondence should be forwarded to H.-L. Choi; Mailing address: 291 Daehak-ro, Rm. E4-C327, C-FRIEND Field Robotics Center, KAIST, Yuseong, Daejeon 305-701, Rep. of Korea.; Tel:+82-42-350-3727; E-mail: hanlimc@kaist.ac.kr}
}


\maketitle


\begin{abstract}
This paper presents a solution procedure of search parameter optimization for minimum load ensuring desired one-off and cumulative probabilities of detection in a multifunction phased array radar.
The key approach is to convert this nonlinear optimization on four search parameters into a scalar optimization on signal-to-noise ratio by a semi-analytic process based on subproblem decomposition.
The efficacy of the proposed solution approach is verified with theoretical analysis and numerical case studies.
\end{abstract}


\IEEEpeerreviewmaketitle

\section{Introduction}\label{sec:intro}



In a phased array radar, a beam is electronically steered and time for the steering is of the order of microseconds \cite{Sko07}.
This capability facilitates the use of it as a multifunction radar that alternately performs tasks of various functions such as search, confirm, track, and fire control.
Since radar resources (such as time and energy) are limited, it is necessary to effectively allocate/distributed these resources to multiple functions and tasks to maximize overall mission performance (see \cite{But98,Gho06,Gop08,Mir06,Win06,Win07,Din08,Bar09,Cha11} for various approaches to multifunction radar resource management).

One key problem in this resource allocation is search optimization that determines parameters of search beam and pattern to achieve (or enhance) search performance while spending minimal (or fixed) amount of radar resources.
Typical measures for search performance are metrics of target detection performance including one-off/cumulative probability of detection and detection ranges, while a search load, defined by the ratio of radar time allotted for search function, is typically used as a metric of temporal resource consumption.
Since minimum-load search allows more temporal resources for higher-priority radar tasks that are usually sensitive to delays, the search optimization is particularly critical to the overall radar mission performance.

The search performance can be influenced by various search beam parameters and the shape of the search beam lattice.
While the choice of lattice types (i.e., triangular versus rectangular) was known not to produce significant difference in performance in terms of overall energy loss \cite{Hah69} and detection probability \cite{Fie93}, selection of beam parameters can make a significant impact on the performance of the radar search function.
Hence, many previous researches have addressed optimization and/or sensitivity studies on these parameters for single-function search radars \cite{Mal63,Hah69,Mat05} and also for multifunction radars \cite{Bil92,Bil97,Zha06,Haf09}.

Most of the previous work utilized numerical or graphical optimization for various search parameters \cite{Mal63,Hah69,Bil92,Bil97}.
Early work in \cite{Mal63} investigated optimal frame time of a search radar which maximized detection range for a given cumulative probability of detection.
Optimal beam spacing in terms of beam shape/packing loss that can be interpreted as required energy consumption was studied in \cite{Hah69}.
The same approach to parameter optimization for search function of phased array radar was introduced in \cite{Bil92}, and false alarm probability, frame time, beam spacing, and duty factor were used as parameters in the design process.
System properties of a radar are adjustable in the design process, thus the objective function of the optimization in \cite{Bil92} was relative required power, instead of target detection performances, guaranteeing specified track initiation range.
In \cite{Bil97}, track initiation range, 50\% detection range, and relative required power with respect to frame time, dwell time, and beam spacing were studied.

However, the approaches described in the previous paragraph cannot be directly applied to the context of resource management for multifunction radars.
For a multifunction radar, differing from a single dedicated function radar, radar resources can be preferentially distributed to higher priority functions than the search function, especially in multi-target tracking or fire control during an engagement.
Thus, constraints/limitations of radar resources available for the search function should be explicitly taken into account.
Recent studies \cite{Zha06,Haf09} were devoted to search parameter optimization with consideration of the concept of radar search load.
In \cite{Haf09}, frame time, signal-to-noise ratio (SNR), and beam overlap angle were optimized to reduce search load while required track initiation range was maintained.
Search load that consists of time and power load, both of which are functions of dwell time and frame time, was presented in \cite{Zha06}, and the parameters were optimized to maximize track initiation range under the constraint of specified search load.
In the same way as the researches in the previous paragraph, numerical optimizations were applied to the both literatures \cite{Zha06,Haf09}.


It should be noted that availability of analytic solutions for search optimization is very limited.
For a missile approach-warning radar system in which full resources are dedicated to the search function, an analytic solution for optimization with a single design variable, dwell time, was presented in \cite{Mat05}.
In the process of deriving the analytic solution, \cite{Mat05}  also confirmed the numerical results in old literature \cite{Mal63} on the generalization of range interval for averaged single-scan probability detection.

In this paper, a semi-analytic optimization procedure of surveillance beam parameters of multifunction phased array radars is presented for minimization of search load while keeping desired one-off and cumulative probabilities of detection.
Beam width, dwell time, beam spacing ratio, and frame time are included in the beam parameters.
The major finding of this work is that the optimization problem can be decomposed into a set of three subproblems, thus the procedure is threefold:
\begin{enumerate}
\item analytical optimization of beam width, dwell time, and beam spacing ratio with a certain level of SNR;
\item root-finding with respect to frame time with a certain level of SNR;
\item one-dimensional numerical optimization of search load along SNR using optimal beam parameters attained in the previous steps.
\end{enumerate}
Since analytic expressions of three beam parameters are available from the first step, the nonlinear constrained optimization with respect to four variables is reduced to a line search problem of a single variable, i.e., SNR, including iterations of one-dimensional root-finding.
Equivalence of this subproblem-decomposition-based approach to the original nonlinear optimization is shown in the process; numerical results verify the efficacy of the proposed method.


A preliminary version of this work was presented in \cite{jan12}, but it addressed only one of the subproblems in this paper.
This paper exclusively includes:
(a) optimization procedure including an additional parameter (i.e. frame time) and a constraint (i.e. cumulative probabilities of detection),
(b) more theoretical analysis on analytic optimization of the first subproblem, and
(c) additional numerical results and a sensitivity study on radar power.

The rest of the paper is organized as follows:
section 2 formulates the search optimization problem by defining the design variables of search parameters, constraints on detection performance, and the cost function of radar load;
in section 3, the decomposition of the optimization problem and respective solutions of the subproblems are presented;
section 4 demonstrates the optimization procedure with numerical results and discusses some findings;
the paper is concluded in section 5.


\section{Problem Description} \label{sec:prob_des}

\subsection{Search Beam Parameters and Radar Load Measure}

\begin{figure}[t]
\centerline{
    \vspace*{-.0in}
    \includegraphics[width=\columnwidth]{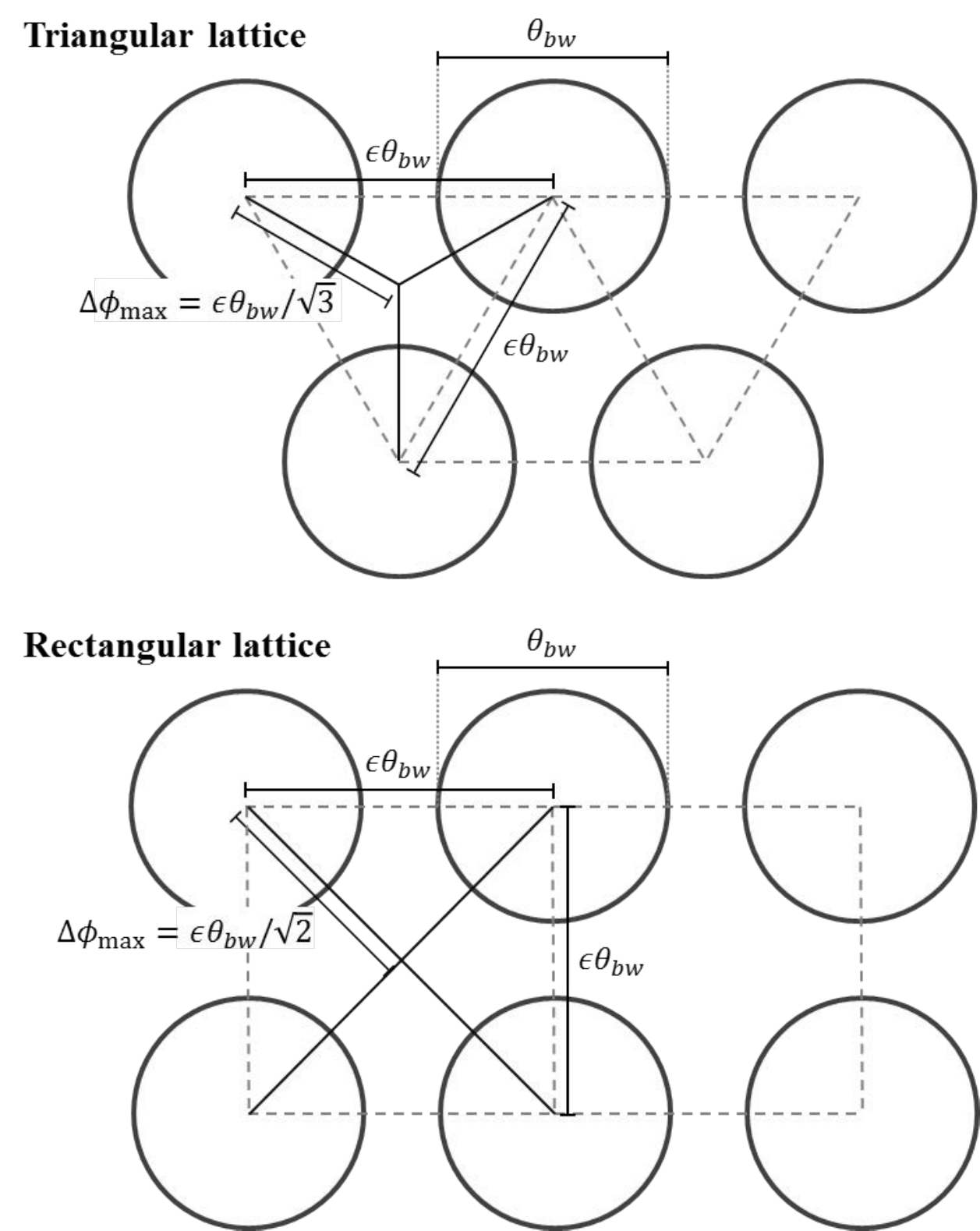}}
    \vspace*{-.0in}
    \caption{The maximum target deviations from beam boresights in triangular/rectangular lattices.}
    \label{fig1}
    \vspace*{-.0in}
\end{figure}

Consider search beam patterns of triangular and rectangular lattices (Fig. \ref{fig1}).
The design variables for the optimization are search beam parameters that decide the pattern and execution of the beams:
\begin{itemize}
\item \textit{Beam Width} ($\theta_{bw}$): The effective width of the radar beam.
\item \textit{Beam Spacing Ratio} ($\epsilon$): Angular distance between adjacent beams normalized by the beam width.
\item \textit{Dwell Time} ($t_d$): Execution time of a search task for a single beam position.
\item \textit{Frame Time} ($t_f$): Period of a search frame, which is a group of beams forming a lattice in an assigned search area.
\end{itemize}

Dimensionless variables associated with the above design variables can also be defined for convenience of analysis:
\begin{eqnarray}
r_{\theta}&:=&\theta_{bw} / \theta_{bw,0} \\
r_{d}&:=&t_d / t_{d,0} \\
r_{f}&:= &t_f /t_{f,0}
\end{eqnarray}
where variables with subscript $0$ denote some reference values.
The beam spacing ratio $\epsilon$ is by definition dimensionless.

Note that a multifunction radar cannot devote the entire temporal resource to surveillance.
Thus, it is important to achieve a certain level of search performance with minimum temporal resource consumption.
The search load is a measure of expected temporal resource consumption by search function:
\begin{itemize}
\item \textit{Search Load} ($L_s$): the proportion of radar time occupied by the execution of search beams in the long-term. 
Thus, if all the search beams have the same dwell time, the search load is the product of the number of beams $N_b$ and the ratio of dwell time $t_d$ to frame time $t_f$:
\begin{eqnarray}
        L_s:=\frac{N_bt_d }{t_f}.
\end{eqnarray}
Let $AZ$ and $EL$ denote effective lengths of search area in azimuth and elevation of the radar, respectively;
the effective length is the length in a plane of radar's direction cosines usually called $UV$-space, but originally $T$-plane \cite{Aul60}.
$q$ is defined as the dimension of the search lattice, i.e., $q=1$ for a one-dimensional and $q=2$ for a two-dimensional lattice.
For the one dimensional lattice $q=1$, one of the effective lengths in azimuth and elevation directions is eliminated or replaced with a non-dimensional property, such as the number of scanning bars.
Then, the number of search beams in the frame required to cover the search area is approximately $AZ\cdot EL/(\theta_{bw}\epsilon)^q$.
Thus, the search load $L_s$ can be expressed in terms of the design variables as:
\begin{eqnarray}
        L_s=\frac{t_d \left( {AZ} \cdot {EL} \right)}{t_f (\theta_{bw} \epsilon)^q}.
\end{eqnarray}
This search load can also be expressed in terms of dimensionless design variables:
    \begin{eqnarray}
        L_s=\eta  r_{d} r_{\theta}^{-q} \epsilon^{-q}   r_{f}^{-1}\label{eq:Ls}.
\end{eqnarray}
with an appropriately defined coefficient $\eta$;
detailed expression of this coefficient is given in (\ref{eq:eta}).
\end{itemize}

\subsection{Search Performance Metrics}

The constraints of the search load optimization are requirements of target detection performance;
one-off probability of detection $P_d$ at  some reference range $R_0$ and cumulative probability of detection $P_c$ at the same reference range are adopted in this work.
The requirements of performance indicate that one-off and cumulative probabilities of detection under some search load have to be larger than or at least equal to respective desired level of probabilities, i.e., $P_{d,des}$ and $P_{c,des}$, that can be written as:
\begin{eqnarray}
P_d \geq P_{d,des}  \label{eq:Pd_req},\\
P_c \geq P_{c,des} \label{eq:Pc_req}.
\end{eqnarray}
The details and expressions of the probabilities of detection are explained in \ref{sec:Pd} and \ref{sec:Pc}, respectively.

\subsubsection{SNR at The Weakest Point of Search Lattice}  \label{sec:snr}

It is known that probability of detection is a function of SNR dependent on the above design variables.
The SNR $S$ of a target that deviates from the beam boresight by an amount of $\Delta \phi$ is:
\begin{equation}
        S (\Delta \phi, R)  =S_0 \left(\frac{R_0}{R}\right)^4\!\left(\frac{\theta_{bw,0}}{\theta_{bw}}\right)^p \frac{t_d}{t_{d,0}}e^{-4a\ln{2}(\Delta\phi/\theta_{bw})^2} \label{eq:S}
\end{equation}
where $S_0$ is the SNR with reference values of range $R_0$, beam width $\theta_{bw,0}$, dwell time of the beam $t_{d,0}$, and zero target deviation $\Delta \phi=0$.
The coefficient $a$ in the exponent is dependent on beam shape loss;
its value is 1 if the radar performs element-level digital beamforming and 2 otherwise.
The impact of beam width on SNR is parameterized by a degree $p$ of beam width ratio.
$p$ is approximately 4 including transmission and reception gains where each gain is a function of a reciprocal of squared beam width \cite{Sko01,Sko07}.

For setting the search performance requirements, the worst-case in a lattice is considered.
With the search lattices in Fig. \ref{fig1}, the weakest point where SNR is the lowest is the point farthest from the centers of beams, i.e., the centroid of the triangle (or the square) with vertices of adjacent beam centers.
The angular distance from a center of a neighboring beam to this centroid can be written as:
\begin{equation}
\Delta \phi_{\max} = k \epsilon \theta_{bw}  \label{eq:del_phi}
\end{equation}
where $k = 1/ \sqrt{3}$ for  a triangular lattice, and $k = 1/\sqrt{2}$ for a rectangular one.

For given beam parameters, the lowest SNR at the reference range $R_0$ where the target deviates from the boresight by $\Delta \phi_{\max}$ is represented as:
\begin{equation}\label{e1}
        S(\Delta \phi_{\max} , R_0)  = S_0 \left(\frac{\theta_{bw,0}}{\theta_{bw}}\right)^p \frac{t_d}{t_{d,0}}e^{-(4a k^2 \ln 2) \epsilon^2 },
\end{equation}
and an associated dimensionless variable can be defined as:
\begin{equation}
        r_S\triangleq \frac{ S(\Delta \phi_{\max} , R_0)}{S_0}  =
       \frac{r_d}{r_{\theta}^{p} }e^{- (4 a k^2 \ln 2) \epsilon^2} \label{eq:rS}.
\end{equation}
Although not explicitly indicated as scripts or arguments, throughout the paper $r_S$ represents dimensionless SNR corresponding to $\Delta \phi = \Delta \phi_{\max}$ and $R = R_0$.

\subsubsection{One-off Probability of Detection Measure ($P_d$)} \label{sec:Pd}

The one-off probability of detection can be determined by an universal equation \cite{Bar05} for Swerling target models \cite{Swe57,Swe60,Swe65};
The detection probability of a target by noncoherently integrated pulses is expressed as a function of single-pulse SNR $S$, false alarm probability $P_{\text{fa}}$, the number pulses or coherent processing intervals $n_{\text{cpi}}$, and the number of independent Rayleigh-distributed samples $n_e$ \cite{Sko07,Bar05}:
\begin{equation}
\begin{split}
&P_d (S, P_{\text{fa}}, n_{\text{cpi}}, n_e)  \\ &~~~=K_m\left(\frac{K_m^{-1}(P_{\text{fa}},2n_{\text{cpi}})-2(n_{\text{cpi}}-n_e)}{(n_{\text{cpi}}/n_e)S+1},2n_e\right),
\end{split}\label{eq:Pd}
\end{equation}
where
\begin{equation}
        K_m(x,d)=1-\frac{1}{2\Gamma(d/2)}\int_{0}^{x}{(t/2)^{d/2-1}e^{-t/2}dt},
\end{equation}
which is the integral of the chi-square distribution, and $K_m^{-1}(p,d)$ is its inverse function.
As it is a cumulative probability distribution function, $K_m(x,d)$ is a monotonically increasing function of $x$ and bounded between 0 and 1; monotonicity also holds for its inverse function $K_m^{-1}(p,d)$. For different Swerling targets, $n_e$ are given:
\begin{equation}
n_e =
\begin{cases}
        1           & \textrm{Swerling I} \\ 
        n_{\text{cpi}}     & \textrm{Swerling II} \\ 
        2           & \textrm{Swerling III} \\ 
        2n_{\text{cpi}}    & \textrm{Swerling IV}.
\end{cases}
\end{equation}
Note that monotonicity of $K_m(x,d)$ with respect to $x$ results in monotonicity of $P_d$ in (\ref{eq:Pd}) with respect to $S$ for given $P_{\text{fa}}$ and $n_{\text{cpi}}$.
By this monotonicity, the constraint of desired one-off probability of detection in (\ref{eq:Pd_req}) can be converted to a constraint of desired SNR:
\begin{equation}
S\geq S_{des}\text{, or equivalently }r_S \geq r_{S, des}. \label{eq:rS_req}
\end{equation}
$S_{des}$ is required SNR  at $R_0$ and $\Delta\phi_{\max}$ providing $P_{d,des}$ for given $P_{\text{fa}}$, $n_{\text{cpi}}$, and a specific target fluctuation model.
$S_{des}$ can be calculated by the inverse function of (\ref{eq:Pd}):
\begin{equation}
S_{des}=\left(\frac{K_m^{-1}(P_{\text{fa}},2n_{\text{cpi}})-2(n_{\text{cpi}}-n_e)}{K_m^{-1}(P_{d,des},2n_e)}-1\right)\frac{n_e}{n_{\text{cpi}}}.
\end{equation}

\subsubsection{Cumulative Probability of Detection Measure ($P_c$) } \label{sec:Pc}

The cumulative detection probability is also used as a measure of search performance.
As described in \cite{Mal63}, this quantity represents the probability that an approaching target is detected at least once by the time it arrives down at some given range $R$;
in this paper, the expression considered in \cite{Mal63, Sko07} is used to define this quantity.
Assuming scan-to-scan independence, the cumulative probability of detection at $R_0$ can be expressed as:
\begin{equation}
        P_c(t_f,r_S)=\frac{1}{V_c t_f}\int_{0}^{V_c t_f} \left\{1-\prod_{i=1}^{m_f} \left(1-P_d(S_{i}) \right) \right\} dr,\label{eq:Pc}
\end{equation}
where $V_c$ is the target's closing velocity, which often can be assumed to be given as a radar design specification.
$S_i$ is the signal-to-noise ratio for the $i$-th latest scan on the target, which can be expressed as:
\begin{equation}
        S_{i} = \frac{ S_0 r_S  R_0^4  }{(R_0+r+(i-1) V_c t_f)^4} \label{eq:Si}
\end{equation}
with $r_S$ in (\ref{eq:rS}), assuming the target's deviation from the beam boresight is $\Delta \phi_{\max}$.
$m_f$, which is the number of scans (or equivalently, number of search frames) on the target before it reaches down to $R_0$, is typically determined such that the one-off probability of detection for the farthest scan, i.e., $P_d( S_{m_f}) $ is sufficiently small \cite{Mal63, Sko07};
this work uses  $10^{-3}$ for this lower limit on $P_d$.

By setting the reference  frame time to be $t_{f,0}=R_0/V_c$, the cumulative probability of detection requirement becomes:
\begin{equation}
\begin{split}
&     P_c(r_f,r_S)  =\frac{1}{r_f}\int_{0}^{r_f}\left\{ 1-\prod_{k=0}^{m_f-1} \left[ 1 - P_d\left( \frac{S_0 r_S}{(1+y+ k r_f)^4} \right) \right]  \right\}dy. \label{eq:Pc_r}
\end{split}
\end{equation}
For given false alarm rate and target fluctuation model, the one-off probability of detection can be calculated by (\ref{eq:Pd}) as a function of $r_S$ and $r_f$, while $r_S$ can be represented in terms of other search beam parameters $r_\theta$, $\epsilon$, and $r_{d}$.

\subsection{Formulation of Search Beam Optimization} \label{sec:formulation}

With the design variables and performance metrics described in the previous sections, the search beam optimization can be formulated as:
\begin{equation}
\min_{r_\theta,\epsilon,r_d,r_f}  L_s (r_\theta,\epsilon,r_d,r_f)  \label{eq:minLs} \\
\end{equation}
\qquad subject to
\begin{eqnarray}
r_S (r_\theta,\epsilon,r_d) &\geq& r_{S, des} \label{eq:rS_des} \\
P_c \left(r_f, r_S (r_\theta,\epsilon,r_d) \right) &\geq& P_{c,des} \label{eq:Pc_des} \\
r_{\theta} &\in& [ r_{\theta, \min}, r_{\theta,\max}]  \triangleq \mathcal{R}_\theta \label{eq:r_theta_bnd}\\
\epsilon &\in& [0, +\infty) \triangleq \mathcal{E} \label{eq:eps_bnd} \\
r_{d} &\in& [ r_{d, \min}, r_{d,\max} ] \triangleq \mathcal{R}_d \label{eq:r_d_bnd}\\
r_{f} &\in& [0, +\infty)  \triangleq \mathcal{R}_f \label{eq:r_f_bnd}
\end{eqnarray}
As in (\ref{eq:minLs}), the objective is to minimize the search load, which can be expressed explicitly as a function of the dimensionless design variables.
Constraints in (\ref{eq:rS_des}) and (\ref{eq:Pc_des}) ensure satisfaction of desired levels of target detection performances, $P_{d,des}$ and $P_{c,des}$ at the reference range $R_0$ and the weakest point in the search lattice.
Constraints in (\ref{eq:r_theta_bnd}) -- (\ref{eq:r_f_bnd}) represent the range of the design variables that are often given by the radar specifications and practical limitations.
The calligraphic variables are defined for notational convenience.

\section{Solution Approach}\label{sec:approach}

One way to solve the search beam optimization given in section \ref{sec:formulation} is to simply implement some general optimization algorithm (such as gradient-based methods, genetic algorithms, and etc.) to the formulation and obtain the optimized values of the design variables.
Such a way should work to provide an (local) optimal solution to the formulation, but it may incur significant computational load and also it is sometimes hard to interpret the physical meaning of the solutions obtained fully relying on numerical optimization tools.
Therefore, this paper tries to take advantage of the underlying functional properties of the optimization problem in order to decompose it into a series of simpler subproblems with some analytical approach as far as possible.
This semi-analytical approach is certainly computationally more efficient, although computation might not be the major issue, and it can provide more insights on the search optimization problem.

From the formulation, observe that:
\begin{enumerate}
\item The objective function $L_s$ can be expressed as:
\begin{equation}
        L_s = \eta \times \widetilde{L}_s (r_\theta,\epsilon,r_d) \times (r_f^{-1}) \label{eq:div_Ls}
\end{equation}
where $\widetilde{L}_s \triangleq r_{\theta}^{-q} r_d \epsilon^{-q}$.
This shows that $L_s$  consists of a product of the function $\widetilde{L}_s$ that does not depend on $r_f$ and another function $r_f^{-1}$ that only depends on $r_f$.
\item The one-off probability detection constraint does not depend on $r_f$.
\item If $r_S$, which is a function of design variables except $r_f$,  is given, the cumulative probability of detection constraint only depends on $r_f$.
\end{enumerate}

Inspired by the above observation, the beam parameter optimization can be decomposed into three subproblems:

\begin{enumerate}

\item[a)] $(r_\theta,\epsilon,r_d)$-{\it subproblem}: For given $r_S \geq r_{S,des}$,
\begin{equation}
\begin{split}
& \qquad \qquad \min_{(r_\theta,\epsilon,r_d) \in \mathcal{R}_\theta \times \mathcal{R}_d \times \mathcal{E}} \widetilde{L}_s (r_\theta,\epsilon,r_d) \\
&\text{subject to } r_S=r_{\theta}^{-p}r_d \exp\left[ - (4 a k^2 \ln 2) \epsilon^2 \right] \label{eq:sub_a}
\end{split}
\end{equation}

\item[b)] $r_f$-{\it subproblem}: For given $r_S \geq r_{S,des}$,
\begin{equation}
\begin{split}
& \qquad \quad \max_{r_f \in \mathcal{R}_f} r_f \\
&\text{subject to }  P_c(r_f,r_S) \geq P_{c,des} \label{eq:sub_b}
\end{split}
\end{equation}

\item[c)] $r_S$-{\it subproblem}:
\begin{equation}
\begin{split}
& \quad\quad\quad\min L_s(r_\theta^*,\epsilon^*,r_d^*,r_f^*)\\
&\text{subject to }  r_S \geq r_{S,des} \label{eq:sub_c}
\end{split}
\end{equation}
where design variables with asterisks are optimal values obtained from the previous subproblems (\ref{eq:sub_a}) and (\ref{eq:sub_b}).

\end{enumerate}

If there is a mechanism to appropriately search over a single scalar variable $r_S$, the beam parameter optimization problem can be solved by iteratively solving two subproblems (\ref{eq:sub_a}) and (\ref{eq:sub_b}) for given candidate $r_S$s.
Any line search algorithm can be adopted as a tool to select an optimal $r_S$ in the last subproblem (\ref{eq:sub_c}).
Moreover, it will be shown that the global optimal solution to the $(r_\theta,\epsilon,r_d)$-subproblem can be obtained analytically, and also that the $r_f$-subproblem boils down to a root-finding problem on a single scalar variable $r_f$.
Therefore, the original optimization problem in \ref{sec:formulation} is reduced to a line search problem of $r_S$ with iterations of root-finding with respect to $r_f$, where the other optimal design variables are provided analytically.

\subsection{Analytic Solution to $(r_\theta,\epsilon,r_d)$-subproblem}\label{sec:subproblem1}

\begin{figure}[t]
\centerline{
    \vspace*{-.0in}
    \includegraphics[width=\columnwidth]{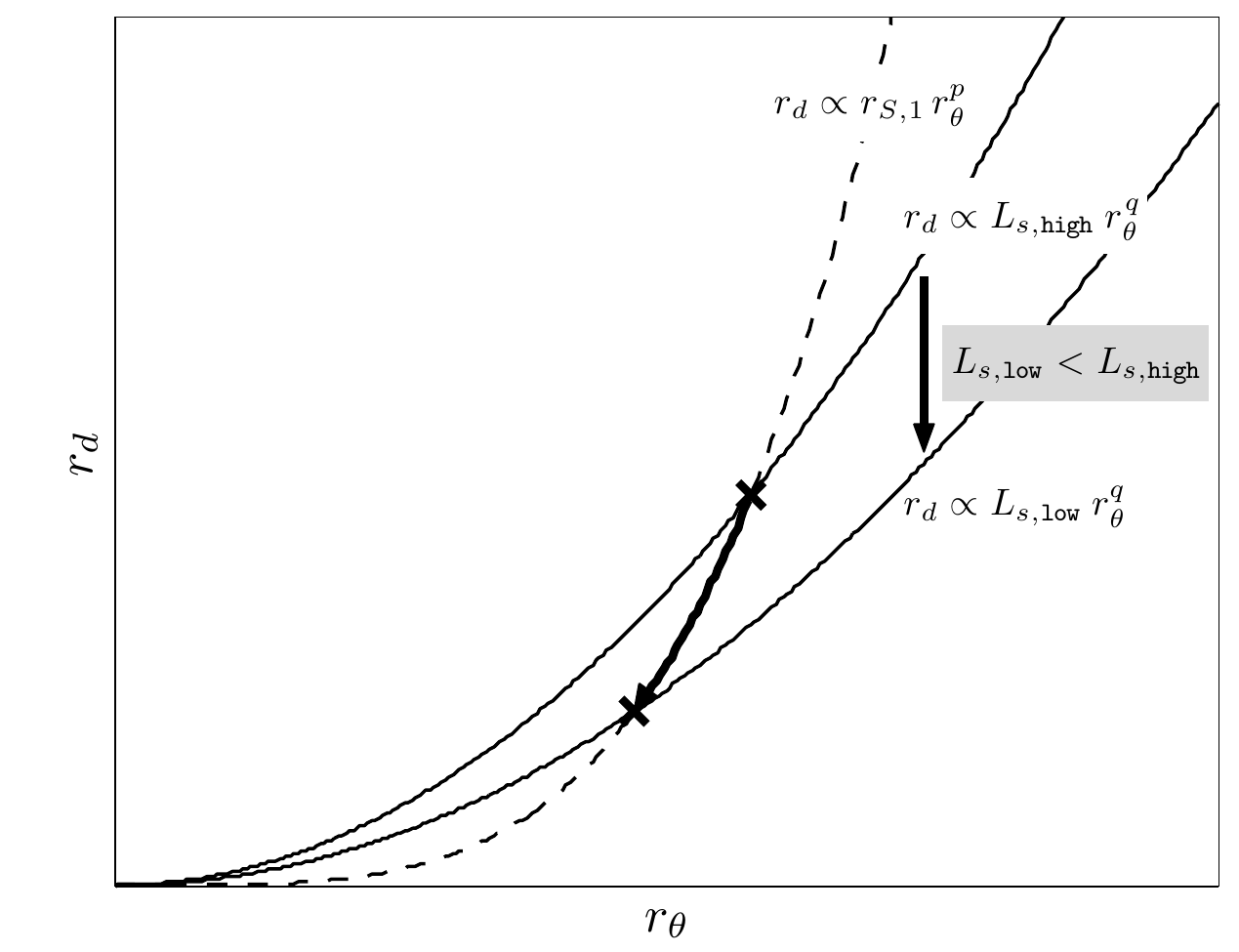}}
    \vspace*{-.0in}
    \caption{The tendency of a feasible $(r_\theta,r_d)$ pair, marked as $\times$, lowering search load where $\epsilon$ and $r_f$ are constant; solid lines are $L_s$ level curves and a dashed line is the level curve for $r_{S,1}$.}
    \label{fig2}
    \vspace*{-.0in}
\end{figure}

\begin{figure}[t]
\centerline{
    \vspace*{-.0in}
    \includegraphics[width=\columnwidth]{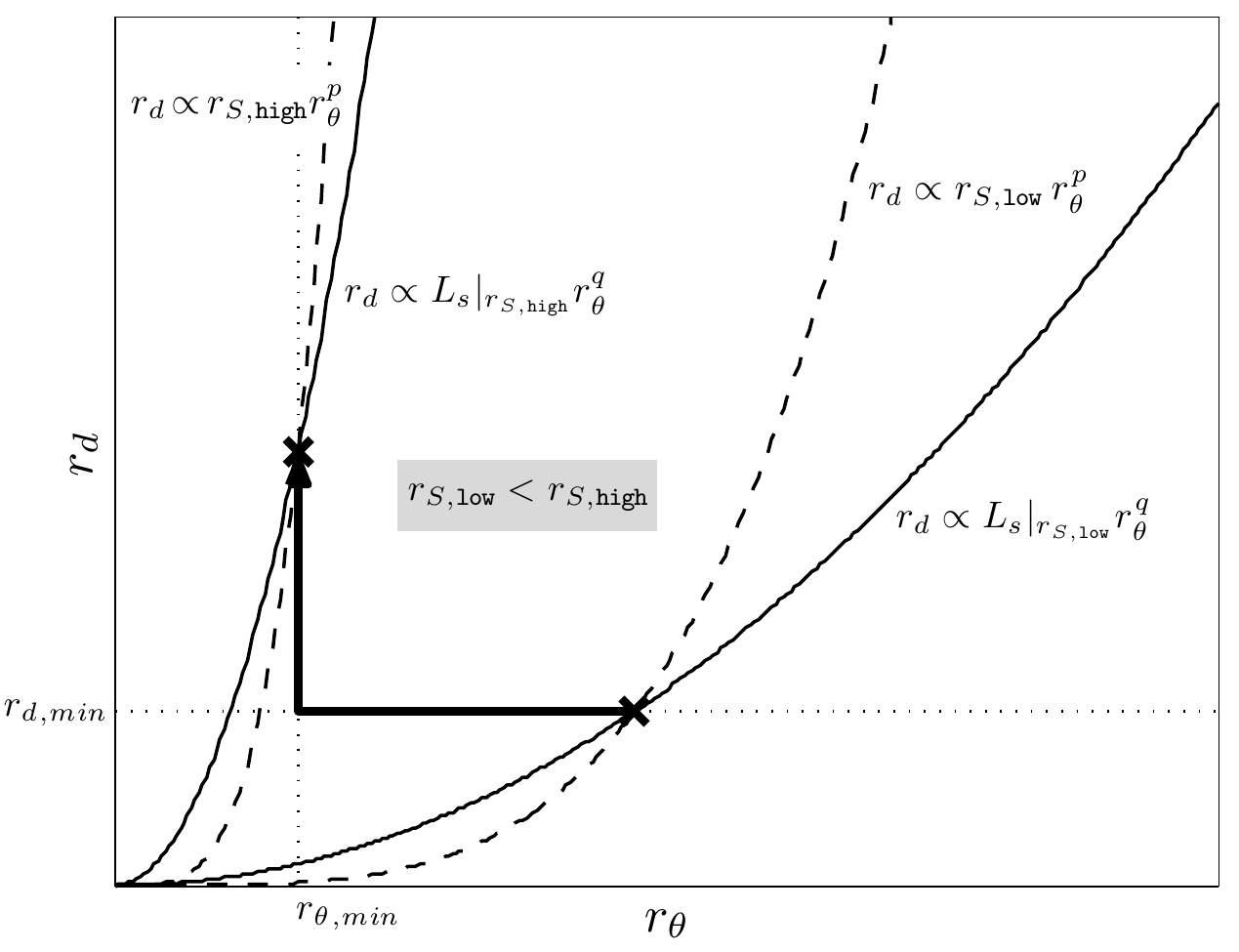}}
    \vspace*{-.0in}
    \caption{The transition of optimal $(r_\theta,r_d)$ pair, marked as $\times$, with limiting bounds where $\epsilon$ and $r_f$ are constant; solid lines are $L_s$ level curves and dashed lines are $r_S$ level curves}
    \label{fig3}
    \vspace*{-.0in}
\end{figure}

Let this section start with some graphical interpretation of the characteristics of the optimal selection of design variables.
From (\ref{eq:Ls}) and (\ref{eq:rS}), if $\epsilon$ and $r_f$ are constant, $r_d$ is proportional to $r_{\theta}^q$ for a given $L_s$ and to $r_{\theta}^p$ for a given $r_S$.
Thus, level curves for the given $L_s$ and $r_S$ can be drawn in the $r_d{-}r_\theta$ plane:
\begin{equation}
        r_d=L_s\frac{r_f}{\eta}(r_\theta\epsilon)^q, \label{eq:lc_Ls}
\end{equation}
\begin{equation}
        r_d=r_Sr_{\theta}^pe^{4ak^2\epsilon^2\ln{2}}. \label{eq:lc_rS}
\end{equation}
Consider two level curves of (\ref{eq:lc_Ls}) for different values of $L_s$, $L_{s,\text{low}}<L_{s,\text{high}}$, and a level curve of (\ref{eq:lc_rS}) for $r_{S,1}$.
Since $p>q$ from the definition of $L_s$ and SNR model in section \ref{sec:prob_des}, the level curve for $r_S$ grows rapidly than the $L_s$ level curves;
and the intersection point of any two level curves for $r_S$ and $L_s$ is the feasible $(r_{\theta},r_d)$ pair for the given $r_S$, $L_s$, $\epsilon$, and $r_f$ (see Fig. \ref{fig2}).
Thus, as depicted in Fig. \ref{fig2}, a lower $L_s$ can be achieved with smaller $r_\theta$ and $r_d$ for the given $r_S$.
This implies that the best feasible solution lies on the lower bounds of $r_\theta$ or $r_d$, i.e., $r_{\theta, \min}$ or $r_{d,\min}$, as otherwise further improvement in the load $L_s$ is possible (see Remark \ref{rem1}).

Now consider level curves for two different $r_S$s,  $r_{S,\text{low}} < r_{S,\text{high}}$, and corresponding optimal $L_s$ level curves and $(r_\theta, r_d)$ pairs (Fig. \ref{fig3}).
Since optimal $(r_\theta, r_d)$ pairs are located only on the lower bounds and the $r_{S,\text{low}}$ level curve lies always below the $r_{S,\text{high}}$ level curve,
$$
r_\theta^*|_{r_{S,\text{low}}} \geq r_\theta^*|_{r_{S,\text{high}}} ,\qquad
r_d^*|_{r_{S,\text{low}}} \leq r_d^*|_{r_{S,\text{high}}}.
$$
Thus, as the desired level of $r_S$ increases, $r_\theta^*$ decreases monotonically and $r_d^*$ increases monotonically.
Fig. \ref{fig3} depicts the transition of the optimal $(r_\theta,r_d)$ pair along $r_{\theta,\min}$ and $r_{d,\min}$.
For very low $r_S$, the optimum is at $r_d^* = r_{d, \min}$ while $r_{\theta}^* > r_{\theta, \min}$.
As $r_S$ increases, the the slope of $L_s$ level curve rises and $r_{\theta}^*$ decreases until it reaches the lower bound $r_{\theta, \min}$.
The optimal pair moves vertically upward along $r_{\theta}^*=r_{\theta,\min}$ after $(r_{\theta,\min},r_{d,\min})$ with a further increase of $r_S$.
Therefore, at least two phases of the optimal point $(r_{\theta}^*,r_d^*)$ transition exist:
a decrease of $r_\theta^*$ with constant $r_d^*$, and then an increase of $r_d^*$ with constant $r_\theta^*$.
For each phase, it will be shown that beam spacing ratio $\epsilon$ remains a constant value.
Since there is a discrepancy between the largest $r_S$ in the first phase and the smallest $r_S$ in the last phase,
beam spacing ratio $\epsilon$ decreases to sustain growing $r_S$ level between the two phases, while $r_\theta^*$ and $r_d^*$ are fixed at their minimum bounds $(r_{\theta,\min},r_{d,\min})$.

Therefore, for varying level of $r_S$, there are three phases in optimal selection of $(r_\theta,\epsilon,r_d)$:

\subsubsection{Phase $0 \rightarrow 1$ (Minimum $r_d$ \& Constant $\epsilon$)} \label{sec:phase01}

When $r_S$ is small enough, $r_d^*$ is fixed at $r_{d,\min}$ and the other two design variables can be determined analytically.
For given $r_S$ and $r_d=r_{d,\min}$, $\widetilde{L}_s$ can be expressed in terms of only $\epsilon$ by eliminating $r_\theta$ using (\ref{eq:rS}):
\begin{equation*}
      \left.  \widetilde{L}_s \right|_{r_{d,\min}} = r_{d,\min}^{1-q/p}r_{S}^{q/p} e^{4a(q/p)k^2\epsilon^2\ln{2}}\epsilon^{-q}.
\end{equation*}
Then, the optimal spacing ratio $\epsilon^*_{0\rightarrow1}$ is determined as
\begin{equation}\label{eq:eps01}
        \epsilon^*_{0\rightarrow1}=k^{-1}(8(a/p)\ln{2})^{-1/2},
\end{equation}
from the stationary point condition:
\begin{equation}\label{eq:partial_Ls_epsilon}
       \left.  \frac{\partial \widetilde{L}_s}{\partial\epsilon}\right|_{r_{d,\min}} = \widetilde{L}_s (8a(q/p)k^2\epsilon\ln{2}-q\epsilon^{-1}) = 0.
\end{equation}
Note that the partial derivative in (\ref{eq:partial_Ls_epsilon}) is a function of only $\epsilon$ (not of $r_\theta$);
thus, the optimal beam spacing ratio is fixed at the constant value $\epsilon^*_{0\rightarrow1}$ (as opposed to the dependency of the optimal beam width).
Once $\epsilon^*_{0\rightarrow1}$ is calculated, the optimal value of $r_{\theta}$ can then be calculated by plugging (\ref{eq:eps01}) into (\ref{eq:rS}).
\begin{equation}\label{eq:rtheta_01}
        r_{\theta,0\rightarrow1}^*=r_S^{-1/p}r_{d,\min}^{1/p}e^{-1/2}.
\end{equation}
The resulting optimal $\widetilde{L}_s$  in this phase becomes:
\begin{equation}\label{eq:Ls01}
        \widetilde{L}_{s,0\rightarrow1}^*=k^q(8e(a/p)\ln{2})^{q/2}r_{d,\min}^{1-q/p} r_S^{q/p}.
\end{equation}
With the transition of $r_{\theta,0\rightarrow1}^*$ from $r_{\theta,\min}$ to $r_{\theta,\max}$, the achievable level of $r_S$ varies from $r_{S,0}=r_{\theta,\max}^{-p}r_{d,\min}e^{-p/2}$ to $r_{S,1}=r_{\theta,\min}^{-p}r_{d,\min}e^{-p/2}$.

\subsubsection{Phase $2 \rightarrow 3$ (Minimum $r_\theta$ \& Constant $\epsilon$)} \label{sec:phase23}

The graphical interpretation suggests that for highest possible  $r_S$, the optimum lies at the lower boundary of $r_\theta$.
 For given $r_S$ and $r_\theta = r_{\theta, \min}$, $\widetilde{L}_s$ can again be written as a function of only $\epsilon$:
$$
\widetilde{L}_s=r_{\theta,\min}^{p-q}r_Se^{4ak^2\epsilon^2\ln{2}}\epsilon^{-q}.
$$
Thus, the stationary point condition
$$
 \frac{\partial \widetilde{L}_s}{\partial\epsilon}=\widetilde{L}_s (8ak^2\epsilon\ln{2}-q\epsilon^{-1}) = 0
$$
leads to constant optimal beam spacing ratio:
\begin{equation}\label{eq:eps23}
 \epsilon^*_{2\rightarrow3}=k^{-1}(8(a/q)\ln{2})^{-1/2}.
\end{equation}
Then, from (\ref{eq:eps23}) and (\ref{eq:rS}) the optimal dimensionless dwell time is obtained as
\begin{equation} \label{eq:rd23}
r_{d,2\rightarrow3}^*=r_{\theta,\min}^pr_Se^{q/2},
\end{equation}
and the resulting optimal $\widetilde{L}_s$ in this phase becomes
\begin{equation}\label{eq:Ls23}
        \widetilde{L}_{s,2\rightarrow3}^*=k^q(8e(a/q)\ln{2})^{q/2}r_{\theta,\min}^{p-q}r_S,
\end{equation}
where the value of $r_S$ in the phase varies from $r_{S,2}=r_{\theta,\min}^{-p} r_{d,\min}e^{-q/2}$ to $r_{S,3}=r_{\theta,\min}^{-p} r_{d,\max}e^{-q/2}$.

\subsubsection{Phase $1 \rightarrow 2$ (Minimum $r_d$ \& Minimum $r_\theta$)} \label{sec:phase12}

Note that $r_{S,2}$ in \ref{sec:phase23} is greater than $r_{S,1}$ in \ref{sec:phase01} with $p > q$, although both correspond to the same $(r_{\theta,\min}, r_{d,\min})$ pair.
As such, in case of $r_S \in (r_{S,1},  r_{S,2})$,  only $\epsilon^*$ can change to accommodate differing levels of $r_S$.
The optimal beam spacing ratio in this phase is obtained as
\begin{equation}\label{eq:eps12}
        \epsilon^*_{1\rightarrow2}=\left(\frac{\ln{(r_S r_{\theta,\min}^p r_{d,\min}^{-1})}}{-4ak^2\ln{2}}\right)^{1/2}
\end{equation}
from
$$
\left. r_S\right|_{r_{\theta,\min}, r_{d,\min}} = \left( r_{\theta,\min}^{-p} r_{d,\min} \right) e^{-4ak^2\epsilon^2\ln{2}}.
$$
Also, the resulting optimal value of $L_s$ is given by
\begin{equation}\label{eq:Ls12}
        \widetilde{L}_{s,1\rightarrow2}^*=r_{d,\min} r_{\theta,\min}^{-q} \left(\frac{-4ak^2\ln{2}}{\ln{(r_Sr_{\theta,\min}^pr_{d,\min}^{-1})}}\right)^{q/2}.
\end{equation}

\subsubsection{Summary}

Through the procedure described in this section, for a given $r_S$ the corresponding phase can be identified and then optimal $(r_\theta,\epsilon,r_d)$ triplet can be obtained analytically.

Also, it is worth noting that:

\begin{Rem} \label{rem1}
If $r_{d,\min} = r_{\theta,\min} = 0$, then the minimum value of $\widetilde{L}_s$ does not exist, but $ \inf \widetilde{L}_s = 0$ with arbitrary small positive values of $r_d$ and $r_\theta$.
\end{Rem}

\begin{Rem} (Convex Program Formulation of $(r_\theta,\epsilon,r_d)$-subproblem):
Since a natural logarithm is a monotonically increasing function, the objective function $\widetilde{L}_s$ can equivalently be replaced by $\log \widetilde{L}_s$
\begin{eqnarray*}
\log \widetilde{L}_s& =& - q \log r_{\theta} + \log r_d - q \log \epsilon \\
& = & - q l_\theta + l_d - q \log \epsilon
\end{eqnarray*}
where $l_\theta = \log r_\theta$, and $l_d = \log r_d$.
Thus, the objective function is a convex function of the new set of design variables $(l_\theta,\epsilon,l_d)$.
By taking the logarithm on both sides of the $r_S$, the requirement constraint then can be written as:
$$
p l_\theta - l_d + (4a k^2 \ln 2) \epsilon^2 +  \log r_S  = 0,
$$
which is a convex constraint on $(l_\theta,\epsilon,l_d)$.
The bounding constraints on $r_\theta$ and $r_d$ can equivalently be written as simple bounding constraints on $l_\theta$ and $l_d$.
Therefore, the problem of minimizing $\log \widetilde{L}_s$ with decision variables of $l_\theta$, $\epsilon$, and $l_d$ is a convex program.

For a convex program, a point satisfying Karush-Kuhn-Tucker (KKT) condition is the global optimum once some mild constraint qualification is satisfied \cite{boyd_cvx}.
Thus, one way to solve the $(r_\theta,\epsilon,r_d)$-subproblem is to pose the equivalent convex program on $(l_\theta,\epsilon,l_d)$ and implement a standard KKT-point finding approach (either analytically or numerically), and recover $(r_\theta,\epsilon,r_d)$ from that solution.
Note that the procedure described in this section is essentially equivalent to applying the KKT condition to analytically find the optimum, which in effect identifies which inequality constraints are active and then applies the stationary point condition.
\end{Rem}

\subsection{$r_f$-subproblem as Root-Finding}\label{sec2b}

The $r_f$-subproblem determines maximum possible $r_f$ for a given $r_S$ under the cumulative detection probability constraint.
The following proposition shows that the cumulative detection probability measure is a monotonically decreasing function of $r_f$ for a given $r_S$.

\begin{Prop}
For a given $r_S$, the cumulative detection probability measure $P_c(r_f,r_S)$ is a monotonically decreasing function of $r_f$.
\begin{proof}
Recall the expression of the cumulative probability of detection:
\begin{equation}
\begin{split}
& P_c(r_f,r_S) \\ &~~ =\frac{1}{r_f}\int_{0}^{r_f}\left\{ 1-\prod_{k=0}^{m_f-1} \left[ 1 - P_d\left( \frac{S_0 r_S}{(1+y+ k r_f)^4} \right) \right]  \right\}dy \\ &~~=\frac{1}{r_f}\int_{0}^{r_f}I(y,r_f)dy.
 \end{split} \label{eq:Pc_prop}
\end{equation}
Note that $P_d(S(y))$ in the integrand is a decreasing function of $y$ since $P_d$ is an increasing function of $S$ from (\ref{eq:Pd}).
Thus, the integrand of (\ref{eq:Pc_prop}), denoted as  $I(y,r_f)$, is a decreasing function of $y$ for a given $r_f$.
For a given $r_S$, the sufficiently small lower limit of $P_d$ that determines $m_f$ corresponds to a certain constant far range, and thus $m_f$ decreases with longer frame time $t_f$, i.e., larger $r_f$.
Now consider two levels of $r_f$, $r_{f,a}\leq r_{f_b}$, then $m_f|_{r_{f,a}}\geq m_f|_{r_{f,b}}$.
Since $P_d\in [0,1]$ and $P_d$ is a decreasing function of $r_f$ for given $y$, larger $r_f$ results in a larger product $\Pi$ in (\ref{eq:Pc_prop}) with smaller $m_f$ and thus smaller $I(y,r_f)$;
in other words,
\begin{equation}
        I(y,r_{f,a})\geq I(y,r_{f,b}).
\end{equation}
Thus,
\begin{equation}
        \int_{0}^{r_{f,a}}I(y,r_{f,a})dy\geq\int_{0}^{r_{f,a}}I(y,r_{f,b})dy. \label{eq:ineq1}
\end{equation}
Recalling that $r_{f,a}\leq r_{f,b}$ and $I(y,r_f)$ is a decreasing function of $y$,
\begin{equation}
        \frac{1}{r_{f,a}}\int_{0}^{r_{f,a}}I(y,r_{f,b})dy\geq\frac{1}{r_{f,b}}\int_{0}^{r_{f,b}}I(y,r_{f,b})dy. \label{eq:ineq2}
\end{equation}
From (\ref{eq:ineq1}) and (\ref{eq:ineq2}),
\begin{equation}
\begin{split}
       \frac{1}{r_{f,a}}\int_{0}^{r_{f,a}}I(y,r_{f,a})dy&\geq\frac{1}{r_{f,b}}\int_{0}^{r_{f,b}}I(y,r_{f,b})dy\\
       \therefore\text{} P_c|_{r_{f,a}}&\geq P_c|_{r_{f,b}}.
\end{split}
\end{equation}
\end{proof}
\end{Prop}

Therefore, the maximum $r_f$ that satisfies $P_c(r_f,r_S) \geq P_{c,des}$ is attained when the constraint is satisfied with an equality:
\begin{equation}
        P_c(r_f,r_S) = P_{c,des}. \label{eq:Pc_equality}
\end{equation}
The solution to (\ref{eq:Pc_equality}) can be obtained by any standard root-finding algorithm (e.g., bisection, Newton-Raphson, etc.).

\subsection{$\eta$ Calculation} \label{sec:eta}

The optimization procedure described thus far does not require information about the coefficient $\eta$, but to interpret the optimized result as the objective value $L_s$, $\eta$ is as important as the optimum value of the decision variables.
From (\ref{eq:Ls}) and the definitions of dimensionless design variables, the search load can be written as:
\begin{equation}
        L_s=\frac{t_{d,0}r_d}{r_fR_0/v_t}\frac{AZ\cdot EL}{(\theta_{bw,0}r_{\theta}\epsilon)^q},
\end{equation}
which leads to
\begin{equation}\label{eq:eta}
       \eta=\frac{t_{d,0}v_t}{R_0\theta_{bw,0}^q}AZ\cdot EL.
\end{equation}
If $r_{d,\min}=1$ and $r_{\theta,\min}=e^{-1/2}$,
\begin{equation}\label{eq:eta_norm}
        \eta = \frac{t_{d,\min}v_t}{R_0\theta_{bw,\min}^qe^{q/2}}AZ\cdot EL.
\end{equation}
In the next section, the reason for the selection of specific values of $r_{d,\min}$ and $r_{\theta,\min}$ will be explained.

\section{Numerical Results}\label{sec:numerics}

\subsection{$(r_\theta,\epsilon,r_d)$-subproblem Results}

In this section, the analytic solutions given in section \ref{sec:subproblem1} are compared with solutions by numerical optimizations.
Hereinafter $r_{\theta,\min}$ and $r_{d,\min}$ are designated to be $e^{-1/2}$ and $1$, respectively, to adjust $r_{S,1}$ to 1;
it can be done by tuning references, i.e., $\theta_{bw,0}$ and $t_{d,0}$ since $\theta_{bw,\min}$ and $t_{d,\min}$ are given by radar specifications.
This is for consistency of analysis and convenience of comparison: this normalization helps to compare simulation results of different parameters, such as $q$, and target fluctuation models, and to apply the optimization results with dimensionless variables to a practical radar.
Analytic solutions with normalized design variables are summarized in Table \ref{tab:summary}.

The bounds of parameters are selected as follows:
$\theta_{bw,\max}/\theta_{bw,\min}=4$, $t_{d,\max}/t_{d,\min}=8$,
thus $r_{\theta,\min}=e^{-1/2}$, $r_{\theta,\max}=4e^{-1/2}$, $r_{d,\min}=1$, and $r_{d,\max}=8$.
For a search beam lattice, triangular one is used, thus $k=1/\sqrt{3}$.
$a$ and $p$ are assumed to be 2 and 4, respectively.
For comparison between analytic and numerical solutions, five equally spaced $r_S$ points are selected for each phase: $[r_{S,0},r_{S,1}]$, $[r_{S,1},r_{S,2}]$, and $[r_{S,2},r_{S,3}]$.

\begin{table*}[t]
\vspace*{-.0in}
    \caption{Summary of Optimal Solutions to $(r_\theta, r_d, \epsilon)$-subproblem (Normalized with $r_{\theta,\min} = e^{-1/2}$ and $r_{d,\min} = 1$) }
    \vspace*{-.1in}
\begin{center}    \setlength{\extrarowheight}{5pt}
    \label{tab:summary}
\begin{tabular}[]{l|l|l}
  \hline\hline
  \IEEEeqnarraymulticol{3}{t}{}\\ [.005in]
Phase Transition Points &  \IEEEeqnarraymulticol{2}{t}{ $r_{S,0}=r_{\theta,\max}^{-p}e^{-p/2}$, \qquad $r_{S,1}=1$, \qquad $r_{S,2}=e^{(p-q)/2}$, \qquad $r_{S,3}=r_{d,\max}e^{(p-q)/2}$}\\
  \IEEEeqnarraymulticol{3}{t}{}\\ [.005in] \hline
  Phase $0\rightarrow 1$ (min. $r_d$ \& const. $\epsilon$) & Phase $1 \rightarrow 2$ (min. $r_\theta$ \& min. $r_d$)  & Phase $2 \rightarrow 3$ (min. $r_d$ \& const. $\epsilon$)\\ [.05in] \hline
  $\epsilon^*_{0\rightarrow1}=k^{-1}(8(a/p)\ln{2})^{-1/2}$              & $\epsilon^*_{1\rightarrow2}=\left(\frac{\ln{r_S}-p/2}{-4ak^2\ln{2}}\right)^{1/2}$             & $\epsilon^*_{2\rightarrow3}=k^{-1}(8(a/q)\ln{2})^{-1/2}$ \\
  $r_{d,0\rightarrow1}^*=1$                                             & $r_{d,1\rightarrow2}^*=1$                                                                     & $r_{d,2\rightarrow3}^*=r_Se^{(q-p)/2}$ \\
  $r_{\theta,0\rightarrow1}^*=r_S^{-1/p}e^{-1/2}$                       & $r_{\theta,1\rightarrow2}^*=e^{-1/2}$                                                         & $r_{\theta,2\rightarrow3}^*=e^{-1/2}$ \\
  $L_{s,0\rightarrow1}^*=bk^q(8e(a/p)\ln{2})^{q/2}r_S^{q/p}r_f^{-1}$    & $L_{s,1\rightarrow2}^*=be^{q/2}\left(\frac{-4ak^2\ln{2}}{\ln{r_S}-p/2}\right)^{q/2}r_f^{-1}$  & $L_{s,2\rightarrow3}^*=bk^q(8e(a/q)\ln{2})^{q/2}e^{(q-p)/2}r_Sr_f^{-1}$ \\ [.1in] \hline\hline
\end{tabular}
\end{center}
\vspace*{-.1in}
\end{table*}

\begin{figure}[t]
\centerline{
    \vspace*{-.0in}
    \includegraphics[width=\columnwidth]{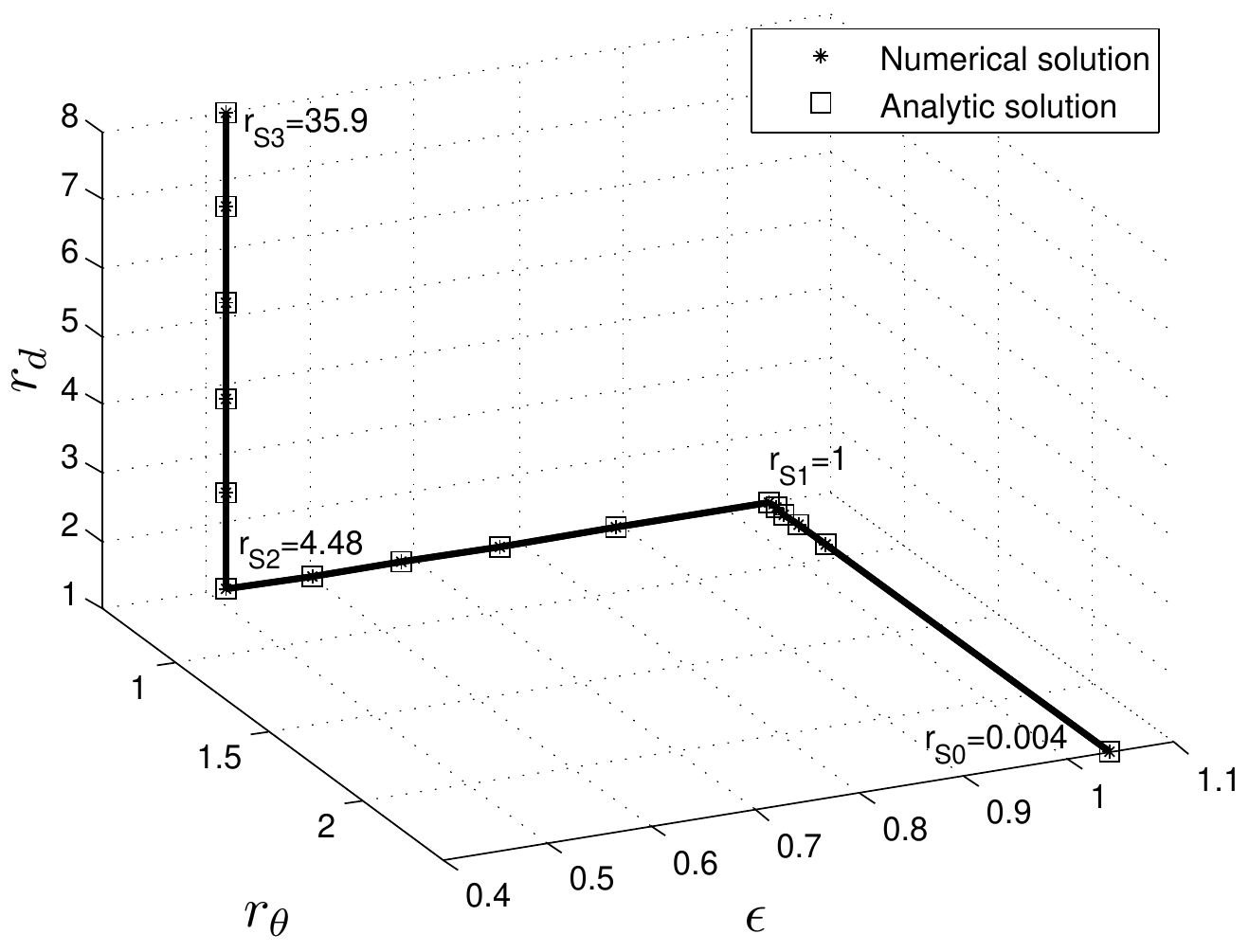}}
    \vspace*{-.0in}
    \caption{Trajectory of optimal parameters $(r_{\theta}^*,\epsilon^*,r_d^*)$ with differing $r_S$ for one-dimensional lattice ($k=1/\sqrt{3}$, $a=2$, $p=4$, $q=1$).}
    \label{fig4}
    \vspace*{.3in}
\centerline{
    \includegraphics[width=\columnwidth]{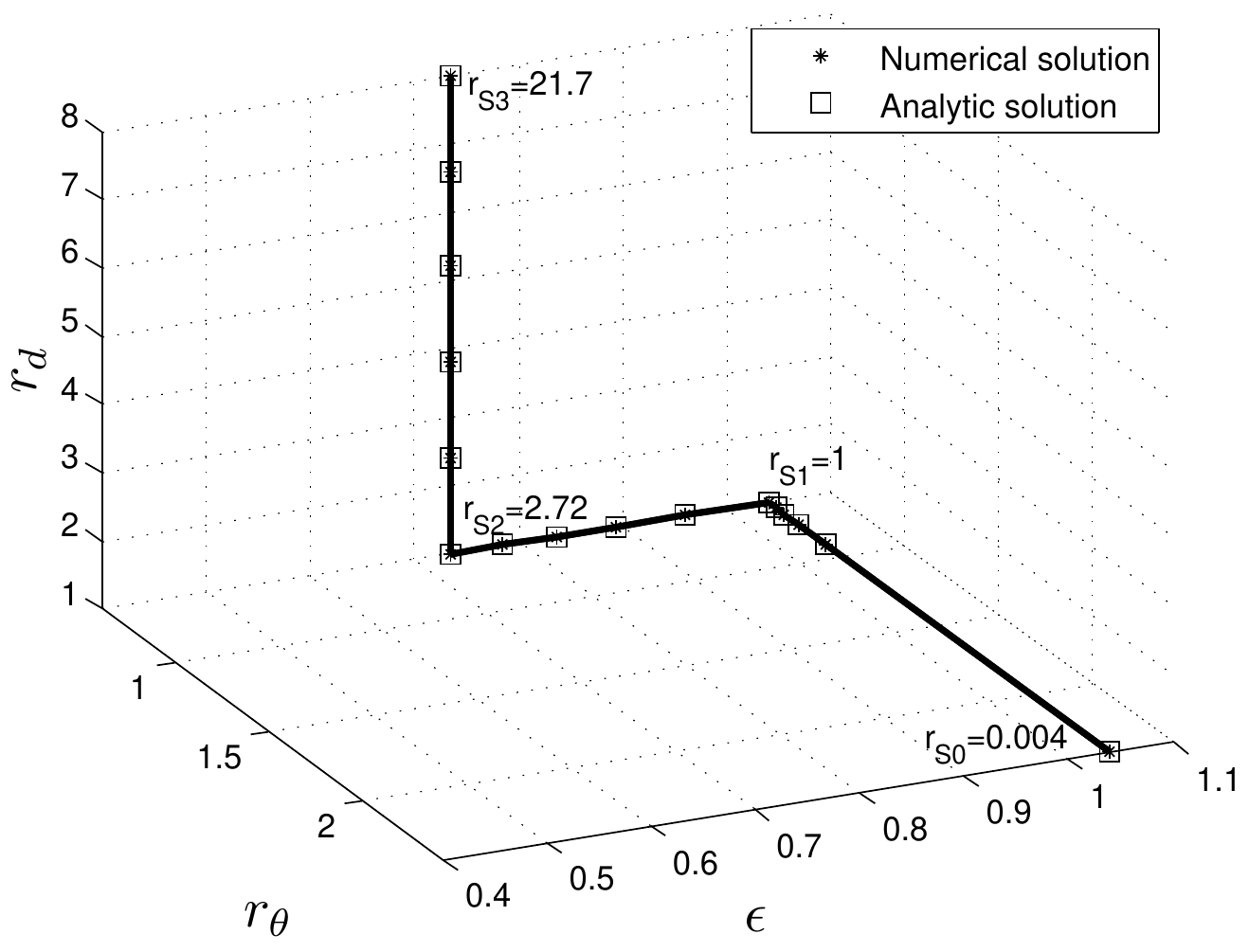}}
    \vspace*{-.0in}
    \caption{Trajectory of optimal parameters $(r_{\theta}^*,\epsilon^*,r_d^*)$ with differing $r_S$ for two-dimensional lattice  ($k=1/\sqrt{3}$, $a=2$, $p=4$, $q=2$).}
    \label{fig5}
    \vspace*{.1in}
\end{figure}

Fig. \ref{fig4} depicts transition of optimal dimensionless parameters $(r_\theta^*,\epsilon^*,r_d^*)$ with respect to $r_S$ when $q=1$.
All analytic solutions are in agreement with the numerical ones.
The transition phases of the optimal solution with respect to the increase of $r_S$ occur in the same order described in section \ref{sec:subproblem1}.
At the early part, $r_S$ increases from $r_{S,0}$ to $r_{S,1}$ while $r_{\theta}^*$ varies from $r_{\theta,\max}$ to $r_{\theta,\min}$ whereas the other two parameters are kept constant.
From $r_{S,1}$ to $r_{S,2}$, $\epsilon^*$ varies from $\sqrt{3/4\ln{2}}$ to $\sqrt{3/16\ln{2}}$.
Afterward, $t_d^*$ increases from $t_{d,\min}$ to $t_{d,\min}$ in proportion to the increase of $r_S$ until $r_{S,3}$ is reached.
The result of the comparison in the case of $q=2$ is presented in Fig. \ref{fig5}.
The overall trend is the same as the one-dimensional case, but $\epsilon^*$ varies from $\sqrt{3/4\ln{2}}$ to $\sqrt{3/8\ln{2}}$.

\begin{figure}[pt]
\centerline{
    \includegraphics[width=\columnwidth]{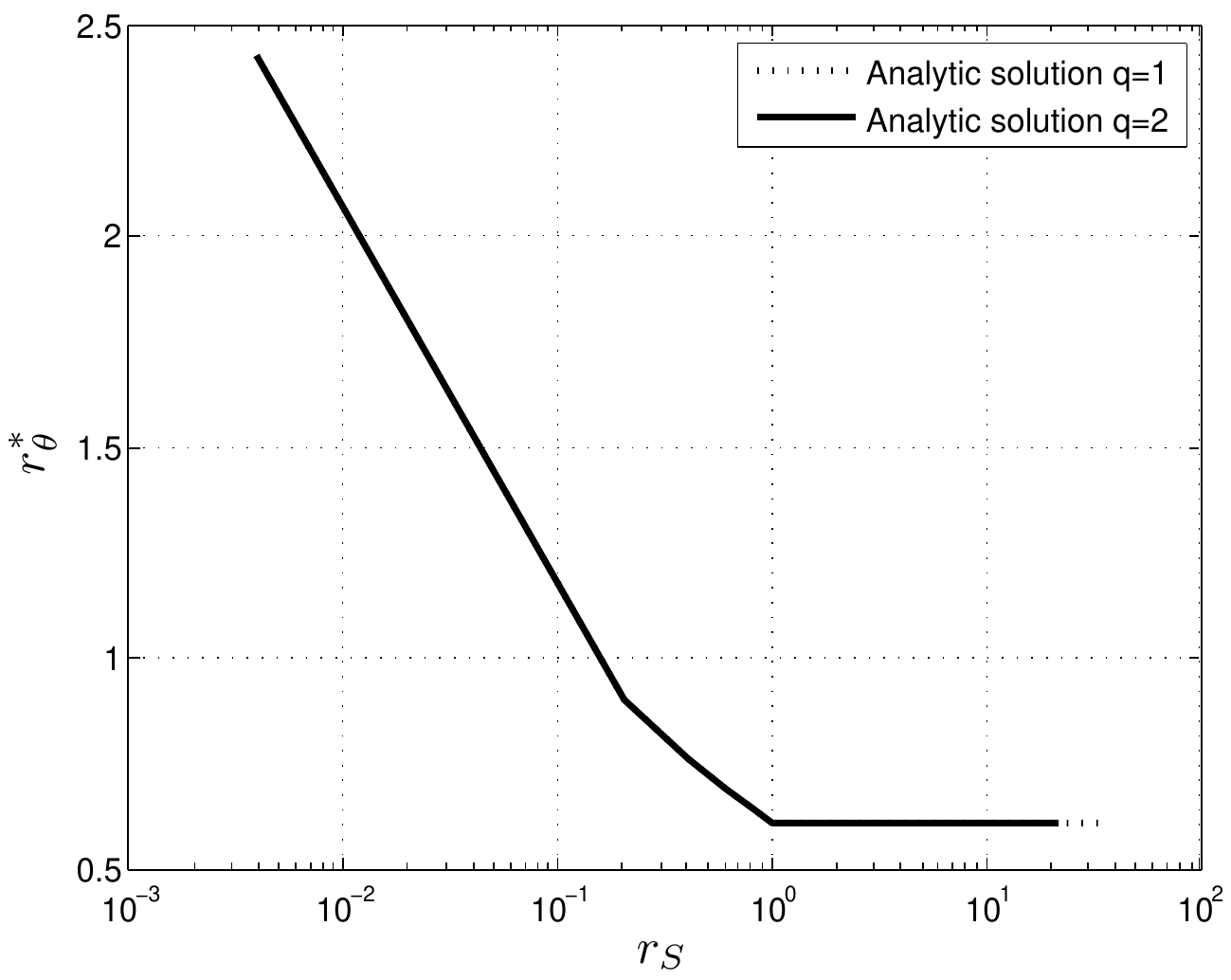}}
    \vspace*{-.1in}
    \caption{$r_{\theta}^*$ for varying $r_S$  (abscissa in log-scale).}
    \label{fig6}
\end{figure}

\begin{figure}[t]
\centerline{
    \includegraphics[width=\columnwidth]{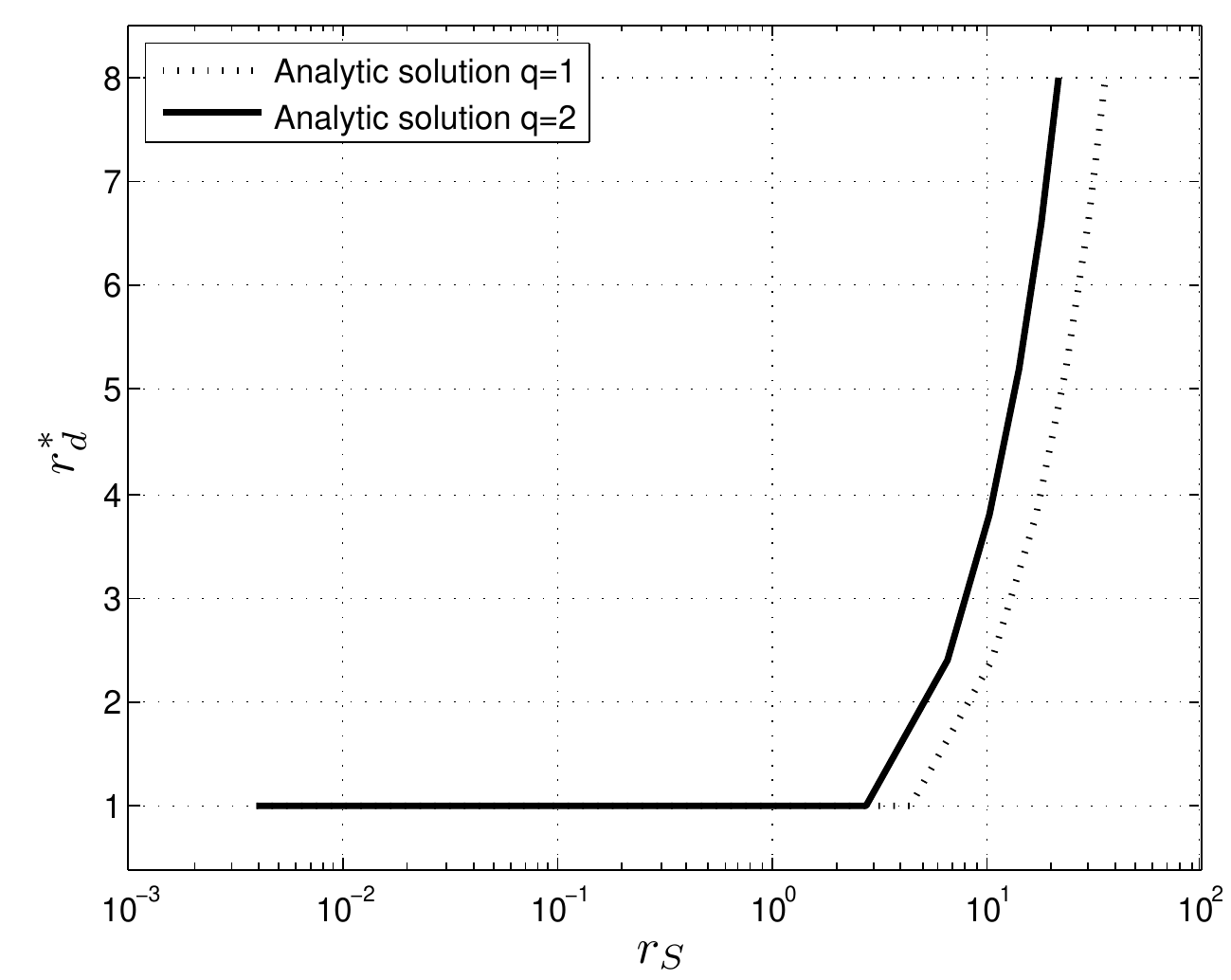}}
    \vspace*{-.1in}
    \caption{$r_d^*$ for varying $r_S$ (abscissa in log-scale).}
    \label{fig7}
\end{figure}

\begin{figure}[t]
\centerline{
    \includegraphics[width=\columnwidth]{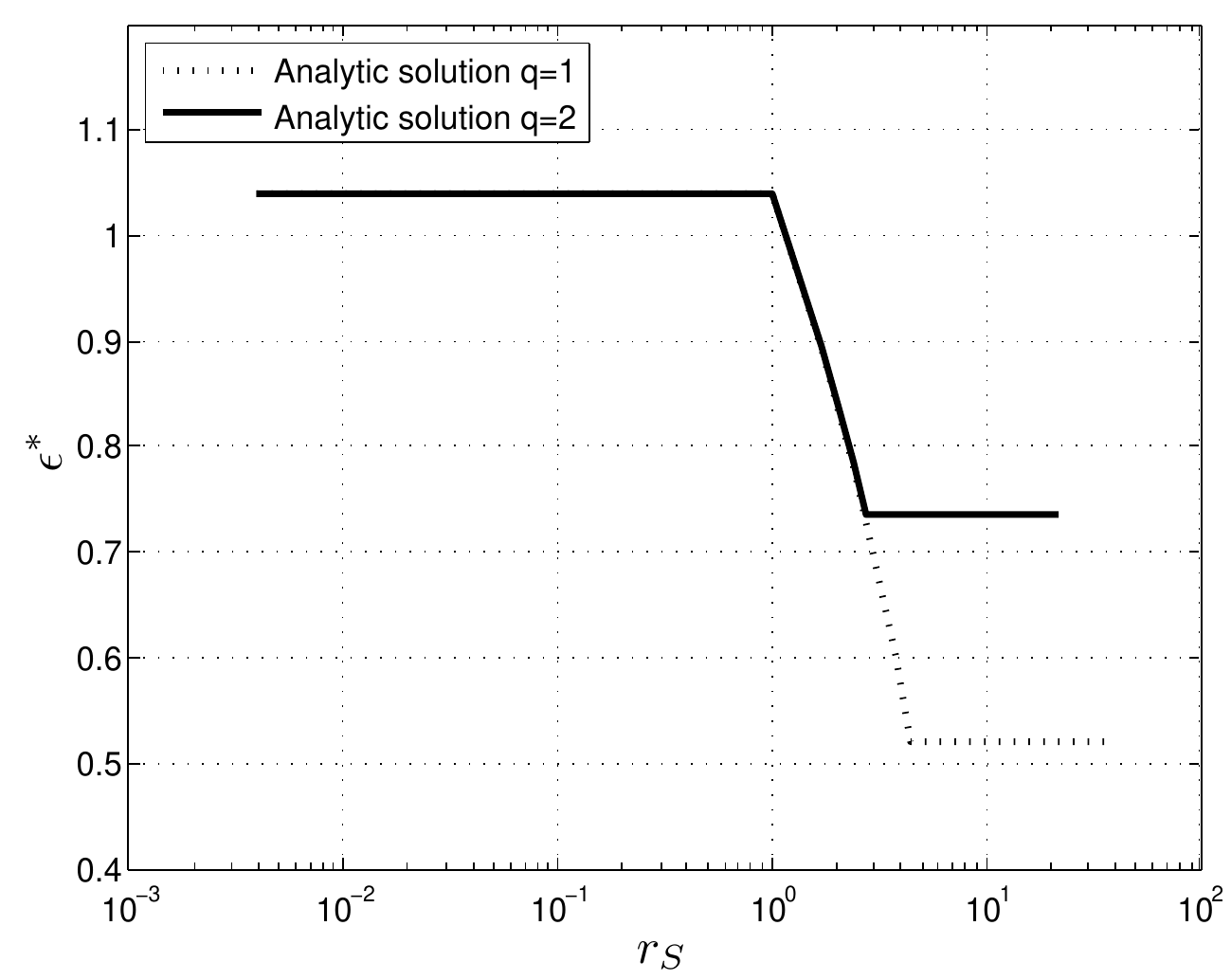}}
    \vspace*{-.1in}
    \caption{$\epsilon^*$ for varying $r_S$ (abscissa in log-scale).}
    \label{fig8}
    \vspace*{.0in}
\end{figure}

The variations of $r_{\theta}^*$, $\epsilon^*$, and $r_d^*$ for two different $q$s are presented in Fig. \ref{fig6} through \ref{fig8}.
Since the analytic optimizations are verified, only those results are plotted.
As $r_S$ increases, $r_{\theta}^*$ starts from the maximum bound and reduces to the minimum bound; finally it becomes saturated.
On the contrary, at first $r_d^*$ remains at the minimum bound and starts to increase after $r_{S,2}$.
$\epsilon^*$ transits from $\epsilon^*_{0\rightarrow1}$ to $\epsilon^*_{2\rightarrow3}$ for each case of $q$.
It should be noted that optimal beam spacing of a search lattice is innately bounded without any constraints of itself.

\begin{figure}[t]
\centerline{
    \includegraphics[width=\columnwidth]{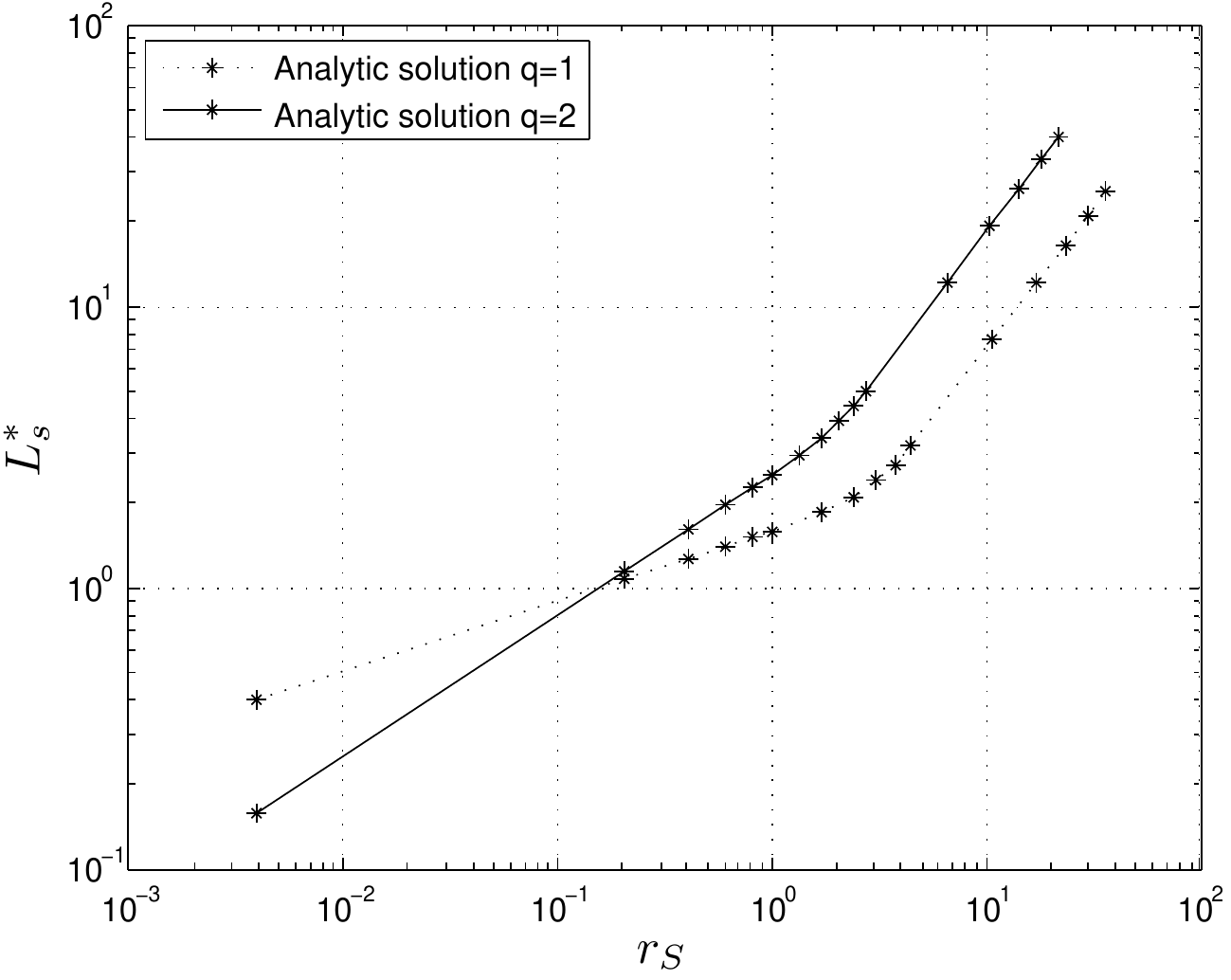}}
    \vspace*{-.0in}
    \caption{Optimal objective value of $(r_{\theta}^*,\epsilon^*,r_d^*)$-subproblem, $\widetilde{L}_s^*$ for varying $r_S$ (log-log scale).}
    \label{fig9}
    \vspace*{-.0in}
\end{figure}

In Fig. \ref{fig9}, the effects of optimal parameters $r_\theta^*$, $\epsilon^*$, $r_d^*$ on the optimal objective value $\widetilde{L}_s^*$ for $\widetilde{L}_s^*$ for varying $r_S$ are presented.
Since the constant $\eta$ can be different among different dimensions of lattices, any comparison of $\widetilde{L}_s^*$ between the results of $q=1$ and $q=2$ is meaningless.
In the log-scales of the both axes, it is concluded that the logarithmic $\widetilde{L}_s^*$ proportionally increases versus the signal-to-noise ratio in dB for $[r_{S,0},r_{S,1}]$;
the slope is increasing in $[r_{S,1},r_{S,2}]$, and is constant again in $[r_{S,2},r_{S,3}]$ with a greater value than in $[r_{S,0},r_{S,1}]$.

\subsection{$r_f$- and $r_S$-subproblem results}

This section describes the calculation process of $r_f$- and $r_S$-subproblem solutions, and explains the search load optimization under practical limits

The optimal search load $L_s^*$ and corresponding $r_S^*$ and $r_f^*$ is calculated as follows.
For each sample point of $r_S$s at intervals of 0.1, the optimal $r_f^*$ satisfying (\ref{eq:Pc_equality}) is obtained by bisection algorithm with  $10^{-4}$ accuracy.
Then, to reduce the effect of uneven solutions from numerical root-finding, $r_f^*$ with respect to $r_S$ is fitted by a 10th order polynomial.
The fitted $r_f^*$ is applied to (\ref{eq:div_Ls}) with $\widetilde{L}_s^*$ from Table \ref{tab:summary} and $\eta$ in (\ref{eq:eta_norm}) to form a $L_s$ curve with respect to $r_S$.
The optimal $L_s^*$ is selected in the $L_s{-}r_S$ curve with $r_S$ satisfying one-off detection probability constraint: $r_S\geq r_{S,des}$.
The optimal parameters $r_\theta^*$, $\epsilon^*$, $r_d^*$, and $r_f^*$ can be computed by the $r_S^*$ of $L_s^*$ using the equations in Table \ref{tab:summary} and the polynomial.

\begin{figure}[h]
\centerline{
    \includegraphics[width=\columnwidth]{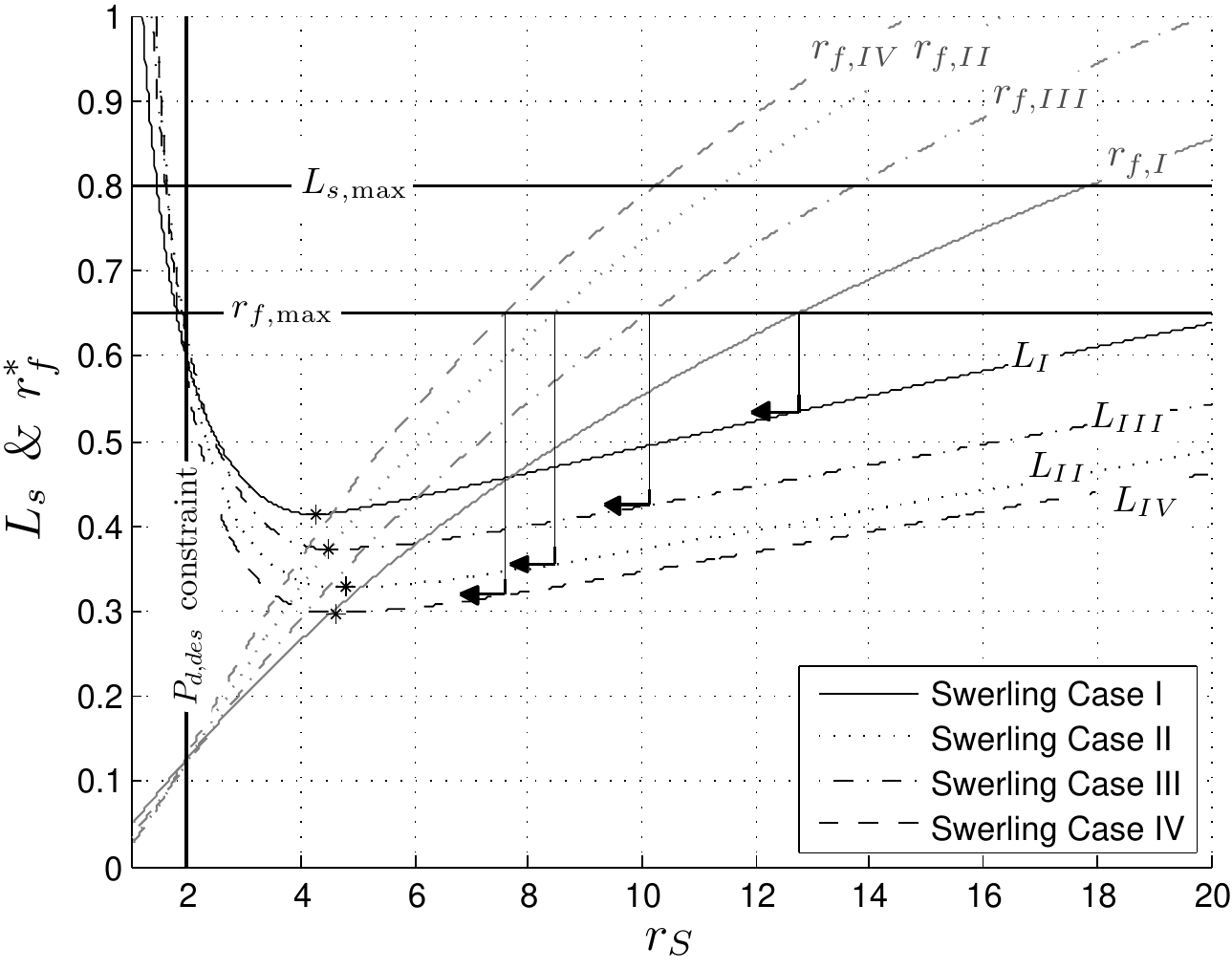}}
    \vspace*{-.1in}
    \caption{$L_s(r_{\theta}^*,\epsilon^*,r_d^*,r_f^*)$ and $r_f^*$ for varing $r_S$ under operational constraints and performance constraint of $P_{d,des}$ where $q=1$; the optimal pair $(r_S^*,L_s^*)$ for each Swerling model is marked as $*$.}
    \label{fig18_a}
    \vspace*{-.0in}
    \vspace*{.2in}
\centerline{
    \includegraphics[width=\columnwidth]{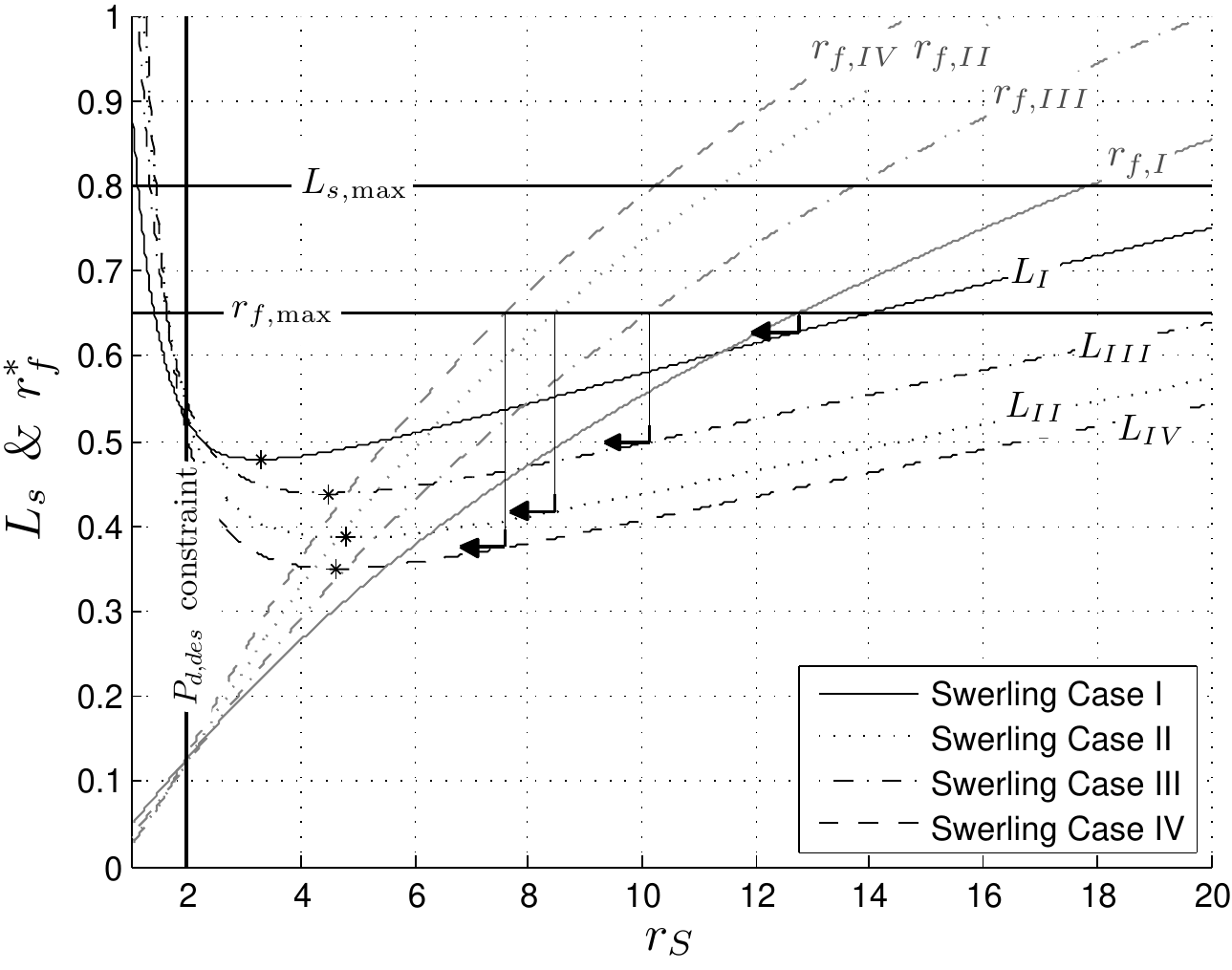}}
    \vspace*{-.1in}
    \caption{$L_s(r_{\theta}^*,\epsilon^*,r_d^*,r_f^*)$ and $r_f^*$ for varing $r_S$ under operational constraints and performance constraint of $P_{d,des}$ where $q=2$; the optimal pair $(r_S^*,L_s^*)$ for each Swerling model is marked as $*$.}
    \label{fig18_b}
    \vspace*{-.0in}
\end{figure}

Fig. \ref{fig18_a} depicts $L_s{-}r_S$ and $r_f{-}r_S$ curves for 4 Swerling cases and the performance constraint of $P_{d,des}$ converted into $r_S$ domain.
For the calculation of the curves, the following is used: $q=1$, $n_{cpi}=4$, $P_{c,des}=0.85$, $t_{d,\min}=5ms$, $\theta_{bw,\min}=2.5^{\circ}$, $R_0=50km$, $v_t=1000m/s$, $AZ=\pm 60^{\circ}$, $EL=16 \text{bars}$, and thus $\eta=0.0385$.
$S_0$ for each target fluctuation model is selected to have the same probability of detection, $P_d(S_0)=0.4$.
The desired level of one-off probability of detection $P_{d,des}$ is assumed to correspond to $r_S=2$ for all target fluctuation models.
The optimal points $(r_S^*,L_s^*)$ for the Swerling models are marked as $*$ in Fig. \ref{fig18_a}.
In the same parametric conditions, the search loads that guarantee specified performances of probability of detection are arranged in descending order as follow: Case I, III, II, and IV.
Fig. \ref{fig18_b} depicts similar results for $q=2$: $EL=\pm 15^{\circ}$, $b=0.0173$ and others are the same with the case of $q=1$.

Additionally, operational conditions that may exist in a multifunction radar are displayed in Fig. \ref{fig18_a}: the maximum affordable level of search load $L_{s,\max}$ and the maximum dimensionless frame time limit $r_{f,\max}$.
These conditions restrict excessive large values of search load and frame time.
$L_{s,\max}$ and $r_{f,\max}$ are assumed to be 0.8 and 0.65, respectively, by a multifunction radar's mission policy.
Since $r_{f,\max}$ restricts the level of $r_f$, the corresponding $r_S(r_{f,\max})=r_{S,\max}$ forms the maximum bound of available $r_S$ (left arrows in Fig. \ref{fig18_a}).
Therefore, the $L_s^*$ is found in $r_S\in[r_{S,des}, r_{S,\max}]$ if the $L_{s,\max}$ constraint is high enough;
with the lower $L_{s,\max}$, the domain of search for the line search problem can be less than $r_S\in[r_{S,des}, r_{S,\max}]$.

\subsection{Effect of Radar Power}

In this paper, the reference SNR $S_0$ can reflect variations in radar supplied power because $S_0$ is SNR at the reference range $R_0$ and the reference conditions.
Since radar power dominantly determines target detection performances and $S_0$ itself can not be normalized, it is worth investigating the effect of radar power on the optimal solutions, especially $r_S^*$ and $r_f^*$.

\begin{figure}[h]
\centerline{
    \includegraphics[width=\columnwidth]{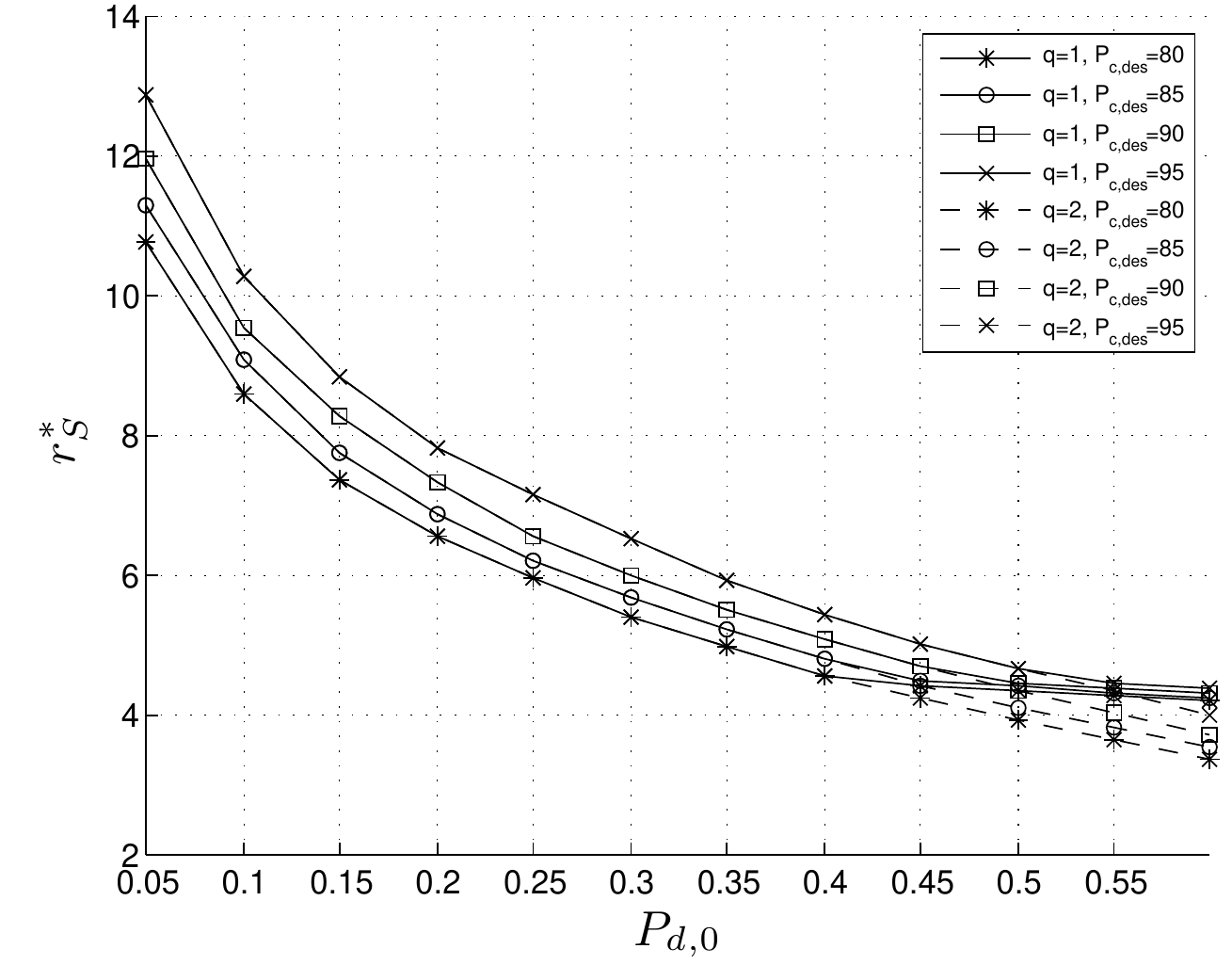}}
    \vspace*{-.1in}
    \caption{$r_S^*$ for differing $P_{d,0}$ (Swerling Case II).}
    \label{fig15}
    \vspace*{.2in}
\centerline{
    \includegraphics[width=\columnwidth]{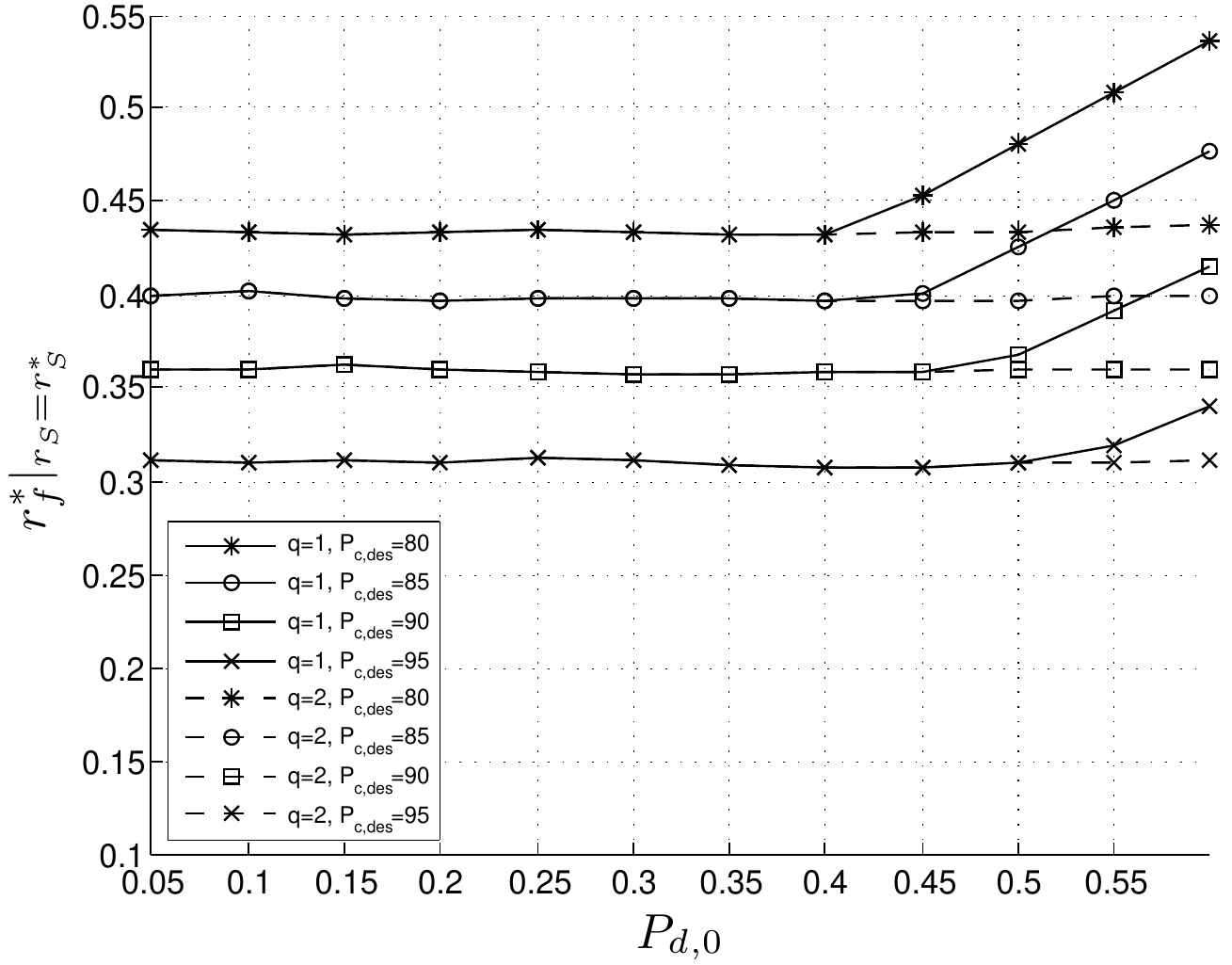}}
    \vspace*{-.1in}
    \caption{$r_f^*|_{r_S=r_S^*}$ for differing $P_{d,0}$ (Swerling Case II).}
    \label{fig11}
    \vspace*{-.0in}
\end{figure}

Consider a reference probability of detection $P_{d,0}{=}P_d(S_0)$ for a specific target fluctuation model, $P_{\text{fa}}$, and $n_{\text{cpi}}$.
Instead of $S_0$, $P_{d,0}$ can be used as a compatible metric of radar power with different target fluctuation models.
Fig.  \ref{fig15} is the graph of $r_S^*$ with respect to $P_{d,0}$ with different $P_{c,des}$s and $q$s for Swerling case II; the constraint of $P_{d,des}$ and the operational conditions are released in the optimization.
$r_S^*$ is greater for larger $P_{c,des}$ at the same $P_{d,0}$ since larger $P_{c,des}$ means a higher level of performance requirement.
As $P_{d,0}$ increases, i.e. $S_0$ increases, $r_S^*$ decreases with diminishing magnitude of slope.
After roughly $r_S=4.48$ for $q=1$ and $r_S=2.72$ for $q=2$, the magnitudes are reduced and the $r_S^*$ curves become linear because $r_S^*$s have transition from $[r_{S,2}, r_{S,3}]$ to $[r_{S,1}, r_{S,2}]$.
It should be noted again that the slope of $\widetilde{L}_s^*$ with respect to $r_S$ is changed in $[r_{S,1}, r_{S,2}]$.

Fig. \ref{fig11} shows $r_f^*|_{r_S=r_S^*}$ curves from the same results of the previous paragraph.
Smaller $r_f^*|_{r_S=r_S^*}$ is needed for larger $P_{c,des}$ at the same $P_{d,0}$, which means faster search is required for higher detection performance.
$r_f^*|_{r_S=r_S^*}$ of each curve remains constant until a certain point of $P_{d,0}$, and it is nearly proportionally increasing with the increase of $P_{d,0}$ after the point.
The points of $r_f^*$-slope transitions are matched with the ones of $r_S^*$ in Fig. \ref{fig15}, i.e. $r_S^*=r_{S,2}$.
$P_c$ is monotonically increasing with $S$ and decreasing with $t_f$, and the decrease of $r_S^*$ with the increase of $P_{d,0}$ is gradual in $r_S^*\leq r_{S,2}$.
Therefore, $r_f^*|_{r_S=r_S^*}$ increases with the increase of $P_{d,0}$ in $r_S^*\leq r_{S,2}$ to satisfies the equality constraint of $P_{c,des}$.
Contrastively, $r_f^*|_{r_S=r_S^*}$ is nearly constant for smaller $P_{d,0}$s since $r_S^*$ decreases with the increase of $P_{d,0}$ in $r_S^*\geq r_{S,2}$.
Although not included in the paper, similar observations can be made for other Swerling Cases.

\section{Concluding Remarks}

The procedure of search load optimization ensuring  desired detection performances for a multifunction radar was studied thus far:
the search load was defined for providing a measure of expected temporal resource consumption by the search function;
for the detection performances, one-off and cumulative detection probabilities for a constant speed incoming target are used;
\iftrue
search beam parameters configuring the search function, i.e. beam width, dwell time, beam spacing ratio, and frame time were selected as design parameters of the optimization.
\else
search beam parameters configuring the search function were selected as design parameters of the optimization: beam width, dwell time, beam spacing ratio, and frame time.
\fi

The formulated optimization problem is decomposable into three subproblems; the first problem can be solved analytically and the second is a root-finding problem.
Since from the first two problems, optimal design variables with respect to the parametric variable, i.e. dimensionless SNR are obtained, the original optimization with four design variables is converted to a line search problem.
Numerical calculations also verified the presented solution approach.

The underlying subject of the search load optimization is multifunction radar resource management with highly requested tasks and functions.
Two radar resource management strategies securing temporal resources for higher priority functions can be suggested:
(1) minimize search load while specific search performances are guaranteed;
(2) maximize search performance with given constant search load.
This paper considered the first strategy, thus any optimization approach for the other strategy is worth a close study.

\section*{Acknowledgments}

This study was supported by the Agency for Defense Development, Korea (Contract Number : UD100057FD).



\newpage
\newpage

\end{document}